\documentclass[twocolumns]{aa}
\usepackage{graphicx}
\usepackage{longtable}
\usepackage{lscape}
\usepackage{txfonts}
\usepackage{natbib}
\begin{document}
   \title{XMM-{\it Newton} spectroscopy of an X-ray selected sample of
RL AGNs
\thanks{Based on observations collected at the Telescopio Nazionale Galileo 
(TNG) and
at the European
Southern Observatory (ESO), La Silla, Chile and on observations
obtained with XMM-{\it Newton}, an ESA science mission
with instruments and contributions directly funded by ESA Member States and 
the USA (NASA)}}
   \subtitle{}
   \author{E. Galbiati\inst{1,2}
	  \and
           A. Caccianiga\inst{1}
	  \and
	  T. Maccacaro\inst{1}
	  \and
	  V. Braito\inst{1}
	  \and
	  R. Della Ceca\inst{1}
	  \and
	  P. Severgnini\inst{1}
	  \and
	  H. Brunner\inst{3}
	  \and
          I. Lehmann\inst{3}
	  \and
	  M.J. Page\inst{4}
}
   \offprints{caccia@brera.mi.astro.it}
   \institute{INAF-Osservatorio Astronomico di Brera, via Brera 28 -
               20121 Milano, Italy\\
	       \email{galbiati, caccia, tommaso, braito, rdc, 
               paola@brera.mi.astro.it} 
	       \and
	       Dipartimento di Fisica, 
              Universit\`a di Milano, Via Celoria 16, I-20133 Milan, Italy
	         \and
		 Max-Planck-Institut f\"ur extraterrestrische Physik, Giessenbachstrasse, 
                 85748 Garching, Germany\\
                 \email{hbrunner@mpe.mpg.de, ile@mpe.mpg.de}
   \and
      Mullard Space Science Laboratory, University College London, 
      Holmbury St. Mary, Dorking, Surrey, RH5 6NT\\
     \email{mjp@mssl.ucl.ac.uk}
                }

   \date{Received 23 July 2004 / Accepted 11 October 2004}

   \abstract{This paper presents the X-ray spectroscopy of an X-ray selected sample
   of 25 radio-loud (RL) AGNs extracted from the XMM-{\it Newton} Bright Serendipitous
   Survey (XBSS). The main goal of the work is to assess and study the origin
   of the X-ray spectral differences usually observed between radio-loud and radio-quiet (RQ)
   AGNs. To this end, a comparison sample of 53 RQ AGNs
   has been also extracted from the same XBSS sample and
   studied together with the sample of RL AGNs.
   Since there are many claims in the literature that RL AGNs have, on average, a flatter spectral
   index when compared to the RQ AGNs, we have focused the analysis on the distribution of 
   the X-ray spectral indices of the power-law component that models the large majority of the 
   spectra in both samples. 
   We find that the mean X-ray energy spectral index is very similar
   in the 2 samples and close to $\alpha_X\sim1$. However, the intrinsic distribution of the
   spectral indices is significantly broader in the sample of RL AGNs. In order to investigate
   the origin of this difference, we have divided the RL AGNs into blazars (i.e. BL Lac objects and
   FSRQs) and ``non-blazars'' (i.e. radiogalaxies and SSRQs), on the basis of the available
   optical and radio information. Although the number of sources is small, we find strong
   evidence that the broad distribution observed in the RL AGN sample is mainly due to the
   presence of the blazars. Furthermore, within the blazar class we have found a link between
   the X-ray spectral index and the value of the radio-to-X-ray spectral index ($\alpha_{RX}$)
   suggesting that the observed X-ray emission is directly connected to the
   emission of the relativistic jet. This trend is not observed among
   the ``non-blazars'' RL AGNs. This favours the hypothesis that, in these latter sources, the
   X-ray emission is not significantly influenced by the jet emission and it has probably an
   origin similar to the RQ AGNs. Overall, the results presented here indicate
   that the observed distribution of the X-ray spectral indices in a given sample of RL AGNs is
   strongly dependent on the amount of relativistic beaming present in the selected sources, i.e.
   on the relative fraction of blazars and ``non-blazars''.
   \keywords{galaxies: nuclei - X-ray: galaxies - BL Lacertae objects: general
               }
   }
   \titlerunning{XMM-{\it Newton} spectroscopy of RL AGNs}
   \maketitle
%

\section{Introduction}

The origin of the differences between radio-loud (RL) and
radio-quiet (RQ) AGNs is one of the fundamental questions in the
AGN phenomenology. The broad and possibly bimodal distribution of
the radio-to-optical flux ratios observed within the class of AGNs
(e.g \citealt{kellermann89}; \citealt{dellaceca94};
\citealt{ivezic02}; but see also
\citealt{white00}; \citealt{cirasuolo03} for evidences against the
bimodality) is not yet fully
understood and it may reflect important differences between the
physical processes at work in the inner regions of the 2 classes
of objects. It has been proposed that the
``radio-loudness" of the AGNs could be directly connected to the
mass of the central black-hole (\citealt{franceschini98}; \citealt{laor00}) or
to its spin (e.g. \citealt{meier99}). Alternatively,
a different accretion rate (e.g. \citealt{woo02}; \citealt{ho02}) or a different
 structure of the accretion disk
(e.g. see \citealt{ballantyne02} and references therein)
may be at the origin of the observed phenomenology.\newline
 In this context, X-ray observations can play a key role  since
they bring information on the innermost region of the AGN
structure.
In particular, a direct comparison between the X-ray properties
of RL and RQ AGNs can shed light on the origin of the ``radio-loudness".

If the main properties of the X-ray spectrum of RQ
AGNs are quite well established, RL AGNs are, in general, less
studied. Early observations carried out with {\it Einstein} have
suggested that RL AGNs have, on average, a flatter spectrum than RQ AGNs
(e.g. \citealt{wilkes87}; \citealt{shastri93}). This difference
is more evident if only RL AGNs with a flat radio spectrum (i.e.
the Flat Spectrum Radio Quasars, FSRQs) are considered, while it
becomes  marginal when only the AGNs with a steep radio spectrum
(i.e. the Steep Spectrum Radio Quasars, SSRQs) are compared to the
RQ AGNs (\citealt{canizares89}). This trend has been subsequently
confirmed with ASCA and BeppoSAX data: on the one hand, the X-ray spectrum of a
sample of RL AGNs (mostly composed by FSRQs) studied by \citet{reeves97} and \citet{reeves00} turned out to be
significantly flatter, on average, when compared to RQ AGNs; on the
other hand, this difference in the X-ray spectral index is not
clearly present in the samples studied by \citet{sambruna-er-m99} and \citet{hasenkopf02}, mostly composed by
radiogalaxies or lobe-dominated quasars.

Apart from the spectral index, other differences in the X-ray
spectra of RL and RQ AGNs include a weaker ``reflection''
hump and Fe I emission line in RL AGNs
(e.g. \citealt{hasenkopf02};
 \citealt{eracleous00};
\citealt{sambruna-er-m99}; \citealt{wozniak98}).
These results have been interpreted as related to a different
structure of the accretion disk in RL AGNs (e.g. \citealt{ballantyne02}).
Moreover, \citet{sambruna-er-m99} have pointed out that among
RL AGNs, the presence of a thermal component (from hot
plasma related to the host galaxy and/or to a cluster of galaxies)
is very common.

Altogether, the data collected so far seem to indicate that there
are significant differences in the X-ray spectrum of RL and RQ
AGNs although the details of these differences depend on 
the radio spectral index of the source. In particular, the
presence of a significant contribution to the X-ray emission coming from
the relativistic jets can play an important role in some RL AGNs (like
the FSRQs).

 Chandra and XMM-{\it Newton} can give a fundamental
contribution to this topic: different components of the X-ray emission
in RL AGNs can be resolved spatially, with Chandra, (e.g. \citealt{gambill03}
or \citealt{schwartz04} for a review) or spectrally, with XMM-{\it Newton}
(e.g. \citealt{brocksopp04},  \citealt{ferrero03}, \citealt{page04}).
In this paper, we present a systematic X-ray spectral analysis of
an X-ray selected sample of 25 RL AGNs extracted
from the XMM-{\it Newton Bright Serendipitous Survey} (XBSS, \citealt{dellaceca04}).
In this sample we have distinguished between the AGNs with flat and steep
radio spectrum in order to establish the importance of the jets in the
X-ray emission of RL objects.
In order to assess how different
the selected RL AGNs are with respect to the RQ AGNs, we have also
selected from the same XBSS a ``comparison'' sample of 53 objects,
mostly composed by RQ AGNs. Since the two samples have a common
origin, the only difference being the radio emission, we expect
that any observed difference could be connected to the different
mechanisms at work in the 2 classes of AGNs.\newline
Throughout this paper H$_{0}$=65 km s$^{-1}$ Mpc$^{-1}$, $\Omega_{\Lambda}$=0.7
and $\Omega_{m}$=0.3 are assumed.

\section{The XBSS sample of RL AGNs}

The XBSS (\citealt{dellaceca04}) consists of two flux-limited samples selected
in the 0.5-4.5 keV and 4.5-7.5 keV energy bands respectively
from the analysis of 237 XMM-{\it Newton} archive images.
The flux limit, in both energy bands, is relatively bright
($\sim$7$\times$10$^{-14}$ erg cm$^{-2}$ s$^{-1}$) so as to allow
an accurate analysis of all the selected sources, both from the
X-ray and the optical point of view. The sample selected in the
0.5-4.5 keV energy band, in particular, contains 389 sources and it
is well suited to extract and to study a sub-sample of RL AGNs.

Given the relatively bright X-ray 
flux limit, about 90\% of the optical counterparts of
the X-ray sources are
brighter than mag=21. At these magnitudes,
the majority of RL AGNs are expected to have a radio flux
at the  $\sim$mJy level at $\sim$GHz frequencies given the
typical radio-to-optical flux ratios of the RL AGNs (see next section).
Therefore,
a positional cross-correlation of the XBSS sample with one of the
existing wide-angle radio surveys (NVSS, \citealt{condon98}, or FIRST,
\citealt{becker95}) allows the selection of most of the RL AGNs present in
the sample.
Although the NVSS survey is less deep ($\sim$2.5 mJy) than the FIRST survey
($\sim$1 mJy), it covers a larger area of sky ($\sim$34,000 Deg$^{2}$)
offering the availabilty of radio data (either a detection or an upper limit)
for $\sim$85\% of the XBSS sources.

We have thus cross-correlated the XBSS sample with the NVSS. In total,
there are 332 sources (out of 389) in the XBSS that fall in the area of 
sky covered by the NVSS. 
Given the
good radio (see \citealt{condon98}) and X-ray (see \citealt{dellaceca04}) positions we have
established that a positional tolerance of 10$\arcsec$ is large enough to
find all the radio counterparts of the X-ray sources except for the objects
that are resolved, in the radio, in two or more components
(e.g. the radio galaxies). The positional cross-correlation with a
tolerance of 10$\arcsec$ has produced
28 X-ray/radio matches. For completeness we have subsequently run an additional
cross-correlation, like the one used to select the sources of the REX survey
(\citealt{caccianiga99}), specifically sensitive to complex morphologies (e.g.
double or triple radio sources). However, no sources of this type have been
found. This is consistent with the results of the REX survey, where
the sources with a complex radio morphology (in the NVSS) represent only $\sim$5\% of 
the total radio/X-ray matches at these frequencies and
flux limits (\citealt{caccianiga99}), which corresponds to $\sim$1 source of
this type expected in the XBSS sample.

In order to estimate the reliability of the radio/X-ray matches, we have
computed the expected number of random matches within 10$\arcsec$ using
the source density given by \citet{condon98}. We expect $\sim$0.4 spurious
matches in the sample which implies that one source, at most, could be a
chance radio/X-ray match. As a further confirmation of this result
we have also performed the same cross-correlation
after having positionally shifted the X-ray catalogue along the declination
so that only chance coincidences are expected. We have repeated 11 times the
cross-correlation between the two catalogues with positional offsets ranging 
from 10 to 110 arcminutes. All these 11 cross-correlations
have produced 0 matches thus confirming the negligible number of
expected spurious matches.\newline

In conclusion, we have found 28 sources, out of the 332 XBSS objects 
included in the NVSS sky coverage, that have a flux density at 1.4 GHz
larger than 2.5 mJy. The list of these 28 X-ray/radio matches is 
reported in Table ~\ref{nvs}.

\subsection{The radio-loudness parameter}

The simple detection of a source in the NVSS catalogue does not necessarily
imply that the object is radio-loud, given the range of magnitudes of the 
selected sources.
At the same time, the NVSS flux limit
is not deep enough to guarantee the detection of all the radio-loud AGNs
present in the XBSS sample. We have thus analysed the radio-loudness
parameter (R) of the 28 sources and the distribution of the upper limits on
R of the 304 XBSS sources not detected in the NVSS in order to pin-point all
RL AGNs actually selected, on the one hand, and to estimate the fraction of
RL AGNs that are lost in the cross-correlation because of the NVSS flux limit,
on the other hand.

\begin{figure}
\centering
\includegraphics[width=8cm]{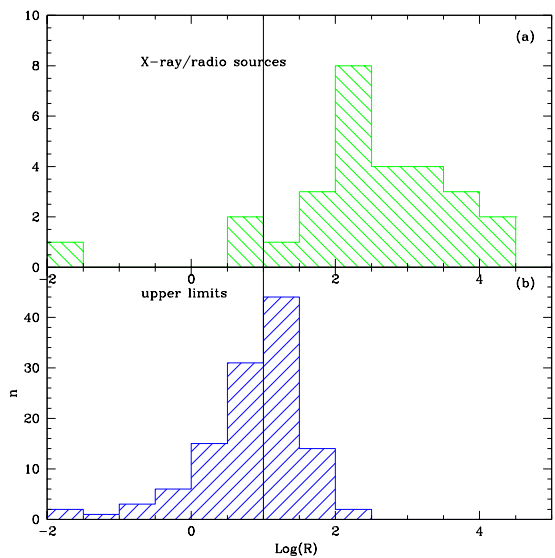}
\caption{Distribution of the radio-loudness parameter (R) for the
28 X-ray/radio matches (panel a) and the upper limits of R for the XBSS
sources not detected in the NVSS (panel b).}
\label{r}
\end{figure}
The parameter R is defined as (\citealt{kellermann89}):
\begin{center}
    R={\Large $\frac{S_{5\,GHz}}{S_{4400\,\AA}}$}
\end{center}
where S$_{5\, GHz}$ is the radio monochromatic flux at 5 GHz and S$_{4400\,\AA}$ is the
optical monochromatic flux at 4400~\AA. 
The radio flux at 5~GHz has been derived from the NVSS flux at 1.4~GHz and K-corrected 
by assuming a mean radio spectral index of 0.7 (S$\propto \nu^{-\alpha}$, \citealt{ciliegi03}).
The optical flux is derived from the optical magnitude taken
from APM (Automatic Plate Measuring machine)
catalogue\footnote{http://www.ast.cam.ac.uk/$\sim$apmcat/} or from the literature and
K-corrected assuming an optical spectral index of 0.5 (\citealt{ciliegi03}).
A magnitude has been obtained for 25 out of 28 sources. The remaining 3 objects are blank-fields
on the DSS material.
For these objects the R parameter is computed assuming a magnitude
equal to the DSS limit (B$\sim$22). In these cases, the actual R parameter is expected to be
larger than the computed value. However, based on what we have found already in the total XBSS sample,
we expect that the optical counterpart in these 3 objects is close to the DSS limit
so that the true value of R should be close to the computed lower limit. In any case, 
these 3 sources have a lower limit on R well above the radio-loud limit (R=10, see below) so that
their classification as RL objects is firm and does not depend on the exact value of
the magnitude.

We finally note that we have used the total magnitude for the computation of the R parameter 
without attempting to separate the contribution due to the host galaxy. The importance
of the correct estimate of the optical nuclear emission in the computation of the radio-loudness
has been often stressed in the literature (e.g. \citealt{ho01}). 
However, this problem is really critical for
samples containing mostly low-luminosity AGN, for which the emission from the host galaxy
is always relevant, if not dominant. This is not the case with the sample discussed here, 
which should be only marginally affected by this problem. In any case, the computed
R parameter for the few objects in which the  host galaxy is important should be more correctly 
regarded as a lower limit.

In Figure~\ref{r} we show the distribution of the R parameter for the 28 X-ray/radio
matches and the distribution of the upper limits on R for the 304 X-ray sources (falling in
the area of sky covered by the NVSS) not detected in the radio band.
Out of 28 radio/X-ray matches, 25 are radio-loud, using a defining limit of R=10.
We stress here that sources with such a large radio-to-optical flux ratio
typically contain  relativistic jets, i.e. an AGN.  
Indeed, based on the available optical, radio or X-ray information, the presence of an AGN
is confirmed, or strongly suggested, in all these 25 objects (see next section).   
For this reason we will use here the term ``RL AGN'' for all the objects with R$>$10.

We have then used the non-parametric method described by \citet{avni80} to
estimate the real distribution of R for the sample XBSS. 
This method provides an analytic solution for
the best estimate of a distribution function of one (binned) independent variable
(the R parameter) taking into account the upper limits, under the assumption that
these are distributed like the detections.
Based on the distribution of R, derived with the \citet{avni80} method, we have
estimated that about 10 RL AGNs present in the XBSS sample have not been detected in the NVSS 
because the
flux limit is not deep enough\footnote{
Indeed, among the XBSS sources that have not been 
detected in the NVSS we
have found one object optically classified as BL Lac, which is expected to belong
to the radio-loud population. Deeper radio observations, reaching a limit of $\sim$0.2 mJy,
should be able to detect all the radio-loud sources in the sample.}. 
Therefore, the expected number of RL AGNs among the 
332 XBSS sources covered by the NVSS data is about 35 (25 detected and $\sim$10 not-detected)
which corresponds to a fraction of 10.5\% of the X-ray sources. 
Since the fraction of AGNs in the XBSS sample is
about 81\% (\citealt{dellaceca04}), the fraction of RL AGNs among the total
population of X-ray selected AGNs at the survey flux limit is $\sim$13\%.
This fraction is consistent with what has been found in previous X-ray surveys, like the EMSS survey 
(\citealt{dellaceca94}), at similar flux limits.

In the rest of the paper we will consider only the sample of RL AGNs, i.e. the one composed
by 25 objects with R$>$10. 

\section{Optical classification}

For the majority (20) of the 25 sources in the sample we have collected a
spectral classification, based either on our own optical spectroscopy or from the literature.
According to the classification criteria described in \citet{dellaceca04}
we have broadly divided the sources into AGN1 (Broad Line AGNs) with broad emission lines
(FWHM$>$1000 km/s) in the optical spectrum, AGN2 (Narrow Line AGNs) showing only
narrow emission lines (FWHM$<$1000 km/s) and objects without strong (EW$<$5\AA) 
emission lines in the spectrum. 

In particular, the AGN1 class includes QSOs, Broad Line radio galaxies,
Sy1 and Narrow Line Seyfert 1 (NLSy1) while the AGN2 class includes type2 QSO, Sy1.8, Sy1.9, Sy2
and Narrow Line radio galaxies. Finally, the class of objects with no emission lines 
includes BL Lac objects and ``normal'' galaxies.

For the classification of the objects with no emission lines in the optical
spectrum, we have used the usual criteria based on the 4000~\AA\ break 
adopted in the major blazar surveys (e.g. the 200 mJy survey, \citealt{marcha96}; 
the RGB survey, \citealt{laurent-muehleisen98}; 
the DXRBS survey, \citealt{perlman98}; \citealt{landt01}; 
the REX survey, \citealt{maccacaro98}, \citealt{caccianiga99}; 
the CLASS survey, \citealt{marcha01}; \citealt{caccianiga01}). 
In particular, we classify a source as  BL Lac object if the 4000~\AA\ break  is absent
(i.e. a completely featureless spectrum) or below 40\%. Sources with a 4000~\AA\ break
larger than 40\% are classified as ``normal'' elliptical galaxies. 
A discussion on the physical meaning of these classification criteria is presented in 
\citet{landt02}.

\begin{figure}
\centering
\includegraphics[width=9cm]{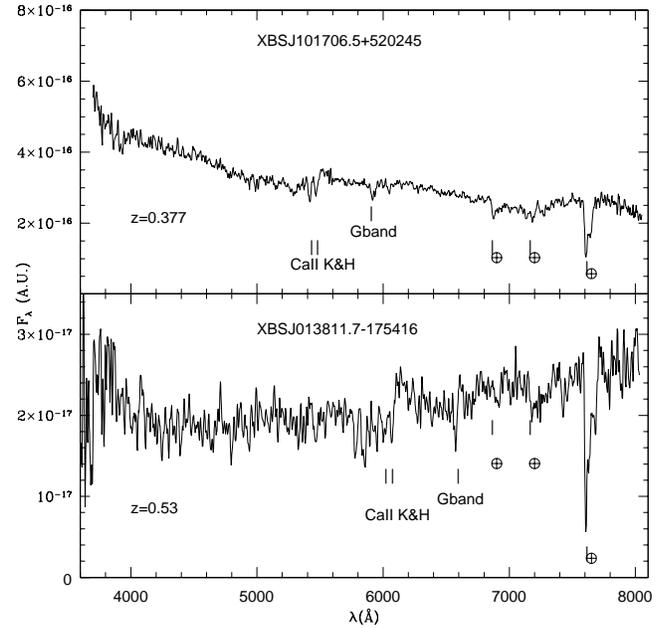}
\caption{Optical spectra of the 2 newly discovered  BL Lacs.}
\label{bllac}
\end{figure}

\begin{figure}
\centering
\includegraphics[width=7cm,angle=-90]{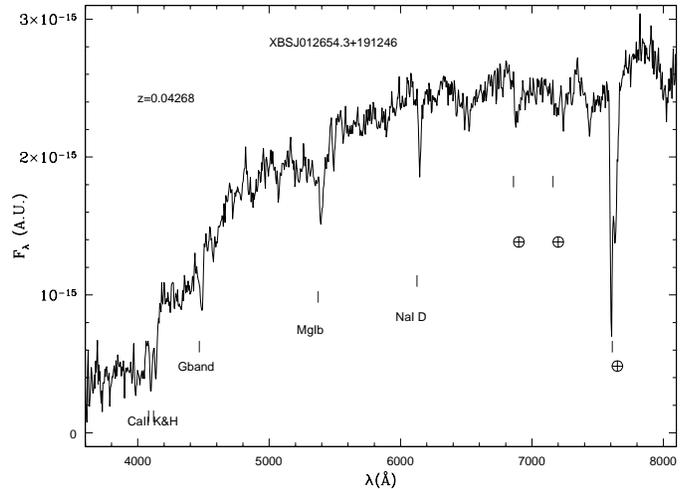}
\caption{Optical spectrum of the source classified as ``normal''
galaxy (a radiogalaxy).}
\label{gal}
\end{figure}

The optical spectra of the 2 newly discovered BL Lacs, taken at the Telescopio Nazionale Galileo 
(TNG) in September 2003 and December 2002, are presented in Figure~\ref{bllac}.
The optical spectrum of the ``normal'' galaxy, taken at the Calar Alto 2.1m telescope in October 2002,
is reported in Figure~\ref{gal}.

In conclusion, the optical spectrum, when available, reveals the presence of an
AGN in 18 out of 20 objects (see Table~\ref{confr}). In the 2 objects, classified respectively as 
``normal galaxy'' and ``cluster of galaxy'', the radio-power is large 
(P$_{1.4 GHz}$=6$\times$10$^{24}$ W Hz$^{-1}$ and 1.6$\times$10$^{25}$ W Hz$^{-1}$ respectively) 
thus confirming, also in these 2 objects, 
the presence of an AGN (i.e. a radiogalaxy isolated or in cluster,
respectively). Similarly to the XBSS sources discussed in \citet{severgnini03}, 
the optical signs of nuclear activity in these 2 radiogalaxies are probably diluted by the 
light of the host galaxy.
Finally, in the 5 sources without an optical spectrum, the presence of an AGN is
inferred by the X-ray spectra (See Section~5). We conclude that, as expected, 
all the sources with R$>$10 are hosting an AGN. 

%

%

    \begin{table*}
\begin{center}
\begin{small}
\caption{The 28 sources resulting from the
X-ray/radio correlation.}
\label{epic}
\begin{tabular}{l l r r r l l r l r r}
    \noalign{\smallskip}
         \hline
name&   CR                       & S$_{1.4}^{int}$ & S$_{1.4}^{pk}$ &
mag & class &redshift& $\alpha_{R}$   & radio class  &Log(R)& cd\\
(1) & (2) & (3) & (4) & (5) & (6) & (7) & (8) & (9) & (10) & (11) \\
    \noalign{\smallskip}
         \hline
\label{nvs}
 \object{XBSJ000100.2$-$250501} & 0.014& 130.0  & 109.9 &  21.8  & AGN1$^1$  & 0.85    &  $>$0.89$^{a}$          & nb  ($\alpha_R$)       & 3.85    &    -          \\
 \object{XBSJ003255.9+394619} &   0.011&  38.8  &  33.6 &  18.7  & AGN1  & 1.14    & 0.58$^{b}$   &    b ($\alpha_R$)             & 2.12    &     -         \\
 \object{XBSJ012505.4+014624} &   0.019& 186.6  & 186.6 &  19.6  & AGN1  & 1.56    & 0.51$^{c}$   &    b ($\alpha_R$; L1)     & 3.25    &         -     \\
 \object{XBSJ012654.3+191246} &   0.010&1408.1  & 947.5 &  15.2$^l$  & G     & 0.04    & 0.53$^{c}$    &   nb (L2)      & 2.10    &       29.80          \\
 \object{XBSJ013240.1$-$133307}& 0.014&   4.6  &   4.6 &  21.4  & AGN2  & 0.56    & $>-$2.06$^{a}$         & ?        & 2.31    &      -       \\
 \object{XBSJ013811.7$-$175416}& 0.031&  15.6  & 15.6  &  20.8  & BL?   & 0.50    & $>-$1.14$^{a}$         &  b (O) & 2.60    &     -         \\
 \object{XBSJ021640.7$-$044404}& 0.026&  88.3  & 88.3  &  16.9  & AGN1  & 0.87    &  0.95$^{d}$       &   nb ($\alpha_R$)            & 1.79    &     -         \\
 \object{XBSJ033226.9$-$274107}& 0.014&  22.5  & 19.6  &  19.2  & AGN1  & 0.73    &  $>-$0.53$^{a}$        &  ?                & 2.05  &    -      \\
 \object{XBSJ052108.5$-$251913}& 0.035&  25.1  & 19.6  &  17.7  & AGN1  & 1.20    &   $>-$0.38$^{a}$       &  ?                    & 1.48    &    -        \\
 \object{XBSJ061342.7+710725} & 0.064&   25.2  & 25.2  & 19.6$^l$  & BL$^2$    & 0.27    &  0.01$^{e}$       &  b (L3; $\alpha_R$)  & 2.31    &            - \\
 \object{XBSJ084026.2+650638} & 0.012&   67.6  & 49.2  & 21.5  & AGN1  & 1.14    & 0.92$^{b}$   &    nb ($\alpha_R$)            & 3.42    &    -        \\
 \object{XBSJ095218.9$-$013643}$^*$& 0.090&   62.2  & 62.2  & 14.4$^l$ & AGN1$^3$  & 0.02    &   $>$0.35$^{a}$         &  RQ        &  0.59  &   49.28     \\
 \object{XBSJ095341.1+014204}$^*$ &  0.163&  9.6  & 9.6   &  17.1$^l$  &   CL  & 0.10    &  $>-$0.95$^{a}$         &  RQ         &  0.85  &  l0.31     \\
 \object{XBSJ101511.8+520708}&   0.017&  77.5  & 77.5  &  21.6  & AGN1  & 0.89    &   0.90$^{b}$ &  nb ($\alpha_R$)          & 3.63    &       8.72      \\
 \object{XBSJ101706.5+520245}&   0.018&  39.7  & 39.7  &  19.9  & BL    & 0.38    &  0.51$^{f}$       &  b (O;$\alpha_R$) & 2.62    &     4.01     \\
 \object{XBSJ102016.1+082143}&   0.011&   3.7  & 3.7  &  $>$ 22.0  & -      & -    &  $>-$1.5$^{a}$         &  ?                        & $>$2.41    &      0.89    \\
 \object{XBSJ111654.8+180304}$^*$ & 0.013&  6.9 &    6.9   & 10.8$^l$  & AGN2$^4$  & 0.003  &  $>-$0.84$^{a}$        &  RQ              &  $-$1.80 &    5.94  \\
 \object{XBSJ111928.5+130250}&   0.018&  40.0  & 14.9  &  18.4  & AGN1  & 2.39    & $>$0.47$^{a}$          &  nb (M)       & 1.80    &      0.44         \\
 \object{XBSJ122658.1+333246}&   0.072&   4.3  & 4.3   & $>$ 22.0  & CL$^5$   & 0.89  & 0.94$^{f}$         &  nb ($\alpha_R$)            & $>$2.53    &      360.00              \\
 \object{XBSJ123538.6+621644}&   0.021&   3.7  & 3.7   &  18.9  & AGN1  & 0.71    &  2.5$^{h}$         &  nb ($\alpha_R$)        & 1.22    &      14.62            \\
 \object{XBSJ124903.6$-$061049}& 0.025&   12.6  & 12.6  & 19.2  & AGN1  & 0.61    & $>-$0.91$^{a}$         &  nb (M)       & 1.88    &      0.59           \\
 \object{XBSJ132105.5+341459}&   0.027& 473.6  & 446.5 &  20.4  &  -     &    -    & 0.82$^{g}$ &nb ($\alpha_R$;M)    & 3.86    &      0.44     \\
 \object{XBSJ133026.6+241520}&   0.021&  46.4  & 41.9  &  20.5  &  -     &    -    &  0.29$^{e}$        &  b ($\alpha_R$)                 & 2.87    &      0.79    \\
 \object{XBSJ133232.6+111220}&   0.011&  26.0  & 26.0  & $>$ 22.0  &       &   -  &  $>$0.06$^{a}$         &  ?                        & $>$3.26    &      8.31     \\
 \object{XBSJ153452.3+013104}&   0.094&1320.4  & 1320.4&  19.8  & AGN1$^6$  & 1.43    & 0.02$^{e}$    &  b ($\alpha_R$; L4)& 4.15   &     24.25 \\
 \object{XBSJ164237.9+030014}&   0.012&  45.5  & 45.5 &   18.6  &  -     &    -    & 0.26$^{e}$    &  b ($\alpha_R$)             & 2.14    &     -      \\
 \object{XBSJ221951.6+120123}&   0.013&   5.8  & 5.8   &  21.3$^j$  &  AGN2 & 0.53    & $>-$1.01$^{a}$         &  ?                        &
 2.36    &    -          \\
 \object{XBSJ235036.9+362204}&   0.024& 316.5  & 273.3 &  19.3  & BL$^7$    & 0.32    &  0.40$^{e}$        &  b (L1,$\alpha_R$)          & 3.22    &    -            \\
        \noalign{\smallskip}
            \hline
\end{tabular}
\end{small}
\end{center}
\footnotesize{
{\bf column 1}: source name (the asterisk indicates that the source is RQ, Log(R)$<$1,
and not considered anymore during the analysis);
\\ {\bf column 2}: 0.5$-$4.5 keV count-rate (counts s$^{-1}$) (see \citealt{dellaceca04} for details);
\\ {\bf column 3}: integrated flux (NVSS) @ 1.4 GHz [mJy];
\\ {\bf column 4}: peak flux density (NVSS)@ 1.4 GHz [mJy/beam] ;
\\ {\bf column 5}: blue magnitude from APM unless specified: $^l$blue magnitude from the literature; $^j$from APM red magnitude and assuming 
the average colour observed for the other sources in the sample (O--E=1.3);
\\ {\bf column 6}: spectroscopic classification (AGN1=type 1 AGN; AGN2=type 2 AGN; G=galaxy; BL=BL Lac; CL=cluster of galaxies). 
Identification and z are from our own spectroscopy except for:

$^1$\citet{fiore03}; $^2$\citet{morris91}; $^3$\citet{nagao01}; $^4$\citet{ho97}; $^5$\citet{ebeling01}; 
$^6$\citet{visvanathan98};$^7$\citet{perlman98};
\\ {\bf column 7}: redshift;
\\ {\bf column 8}: radio spectral index or lower limit (S$\propto \nu^{-\alpha_{R}}$)
\\a= calculated from 1.4 GHz and an upper limit for the flux @ 5 GHz;
\\b= is the mean of the values calculated between 1.4 GHz and other 2 frequencies (5 GHz and 325 MHz);
\\c= is the mean of the values calculated between 1.4 GHz and other 2 frequencies (5 GHz and 365 MHz);
\\d= calculated between 365 MHz and 1.4 GHz;
\\e= calculated between 1.4 GHz and 5 GHz;
\\f= calculated between 325 MHz and 1.4 GHz;
\\g= is the mean of the values calculated between 1.4 GHz and other 3 frequencies (5 GHz, 325 MHz and 365 MHz);
\\h= calculated between 1.4 GHz and 8.5 GHz;
\\{\bf column 9}: radio classification (b= blazar, nb= non blazar, RQ=radio quiet).
\\``$\alpha_{R}$'': from the radio spectral index;
\\``M'': from the radio morphology;
\\``O'': from the optical classification;
\\``L'': from the literature, i.e. 1=\citet{perlman98}, 2=\citet{owen97}, 3=\citet{morris91}, 4=\citet{visvanathan98};
\\{\bf column 10}: radio loudness parameter;
\\{\bf column 11}: core dominance parameter.}
\end{table*}

\normalsize

\begin{table*}
\begin{center}
\caption{Optical properties of the radio loud and of the comparison sample.}
\label{confr}
\begin{tabular}{ l c r c r}
 \noalign{\smallskip}
    \hline
    \hline
 Sample & radio loud & &comparison & \\
\noalign{\smallskip}
    \hline
    \hline
 Number of sources & 25 & &53& \\
\vspace{0.5cm}
Number of identified sources & 20 &82\% & 51 & 96\%\\
Type 1 AGNs & 12 &60\%& 40 &78\% \\
Type 2 AGNs &2 &10\%& 7 &14\%  \\
Galaxies  & 1 &5\% & 2 &4\%\\
Clusters of galaxies & 1 &5\% & 1 & 2\% \\
BL Lacs & 4 &20\% & 1& 2\% \\

    \noalign{\smallskip}
    \hline
    \hline
\end{tabular}
\end{center}
\end{table*}

\section{Radio classification}

In order to further characterize the sources we have collected
all the radio data available from the literature. The main goal is
to distinguish between core-dominated, flat spectrum
objects (e.g. blazars) and lobe-dominated, steep spectrum sources
(e.g. radio-galaxies and Steep Spectrum Radio Quasars, SSRQs). There are indications
in the literature, in fact, that the emission observed in the X-ray band
can be originated in different regions of the AGN depending on the radio
classification of the object: in the case of blazars the X-ray emission is probably
originated from the inner part of the relativistic jet, while, in the
SSRQs and in some radiogalaxies the origin of the X-ray emission is probably
more similar to radio-quiet AGNs (QSOs and Seyfert galaxies,
respectively).
The distinction between blazars and non-blazars would require
specific data in order to measure, for instance, the variability and
the polarisation that are characteristic features of the blazar class.
However, we can make a first attempt at selecting
the blazars in the sample by using the existing radio data taken from  many
different surveys.
Therefore, we have derived (non simultaneous) data at different radio 
frequencies from 
that of the NVSS (1.4 GHz), and/or radio data with a better resolution (i.e. the
data from FIRST survey, made with VLA in B-array) in order to constrain the
core-dominance.

\subsection{Radio spectral indices}
Most of the radio fluxes are collected at 5 GHz from the GB6 catalogue (\citealt{gregory96})
while a few objects have been detected at low frequencies in the WENSS survey
(\citealt{rengelink97}) or in the TEXAS survey (\citealt{douglas96}). Finally,
a few other radio fluxes are taken from the literature.
In total, we have fluxes at one or more frequencies other than 1.4 GHz
for 15 objects out of 25. The typical definition of ``flat'' spectrum radio source
usually requires a spectral index flatter than 0.5 (e.g. \citealt{brunner94}). However, 
when non-simultaneous data, and/or data taken with instruments with different resolutions, 
are used, a more correct limit for the distinction between blazars and
non-blazars is $\alpha$=0.7 (e.g. \citealt{perlman98}; \citealt{cavallotti04}).
By adopting this limit the reliability of the blazar classification
is not 100\%. However, as discussed in \citet{perlman98}, this limit,
coupled with the X-ray emission, leads to a small fraction ($\sim$5\%) of non-blazars wrongly
classified as ``blazars''. Indeed, we have found one object (XBSJ012654+191246), a brigh
($\sim$1.4 Jy) radiogalaxy, whose high resolution radio map shows a clear extended structure
(\citealt{owen97}) in spite of a relatively flat radio spectral index (0.53). In the
following analysis, we consider this object as non-blazar.

\subsection{Radio morphology}
\begin{figure}
\centering
\includegraphics[width=6cm]{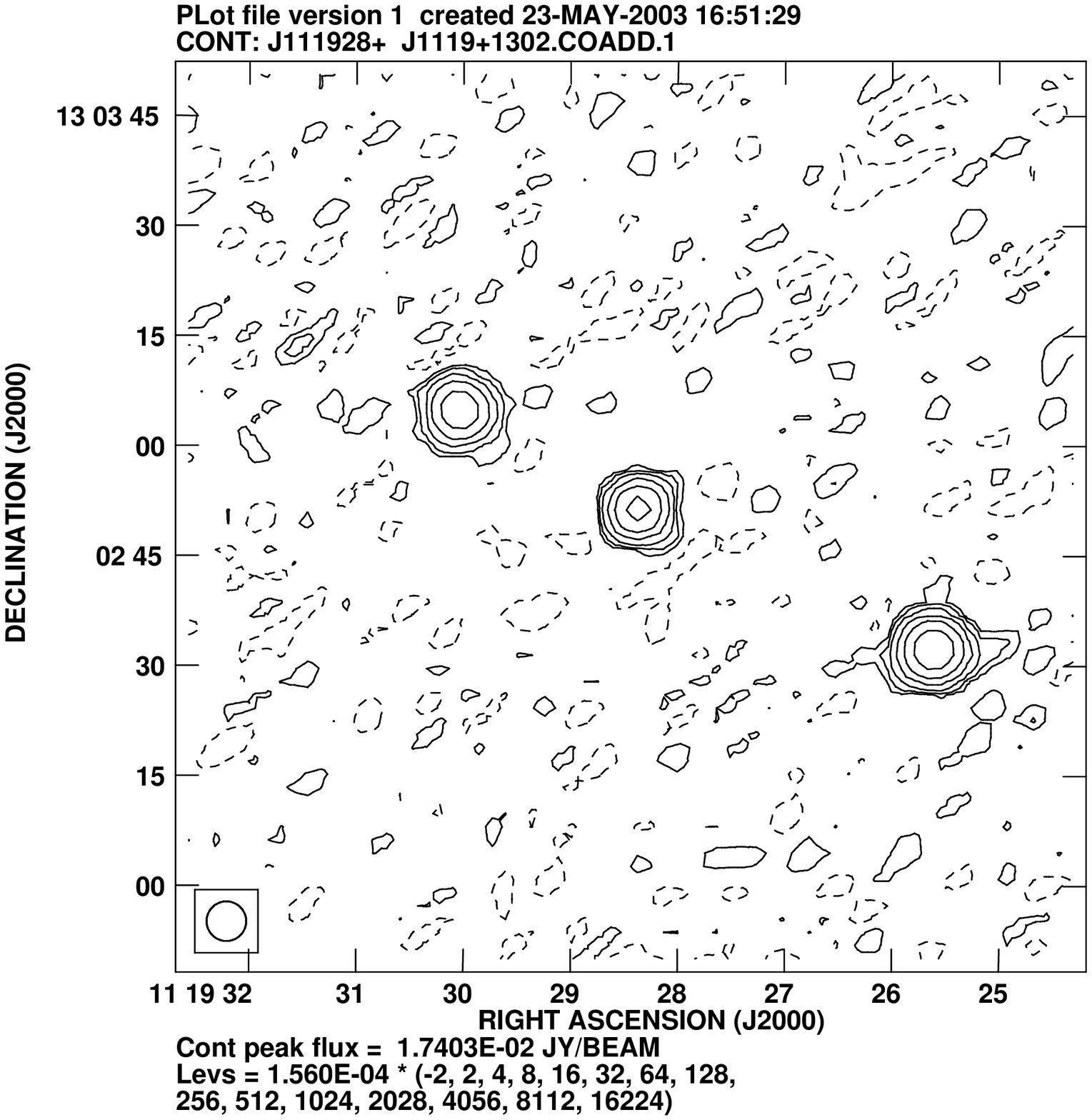}
\includegraphics[width=6cm]{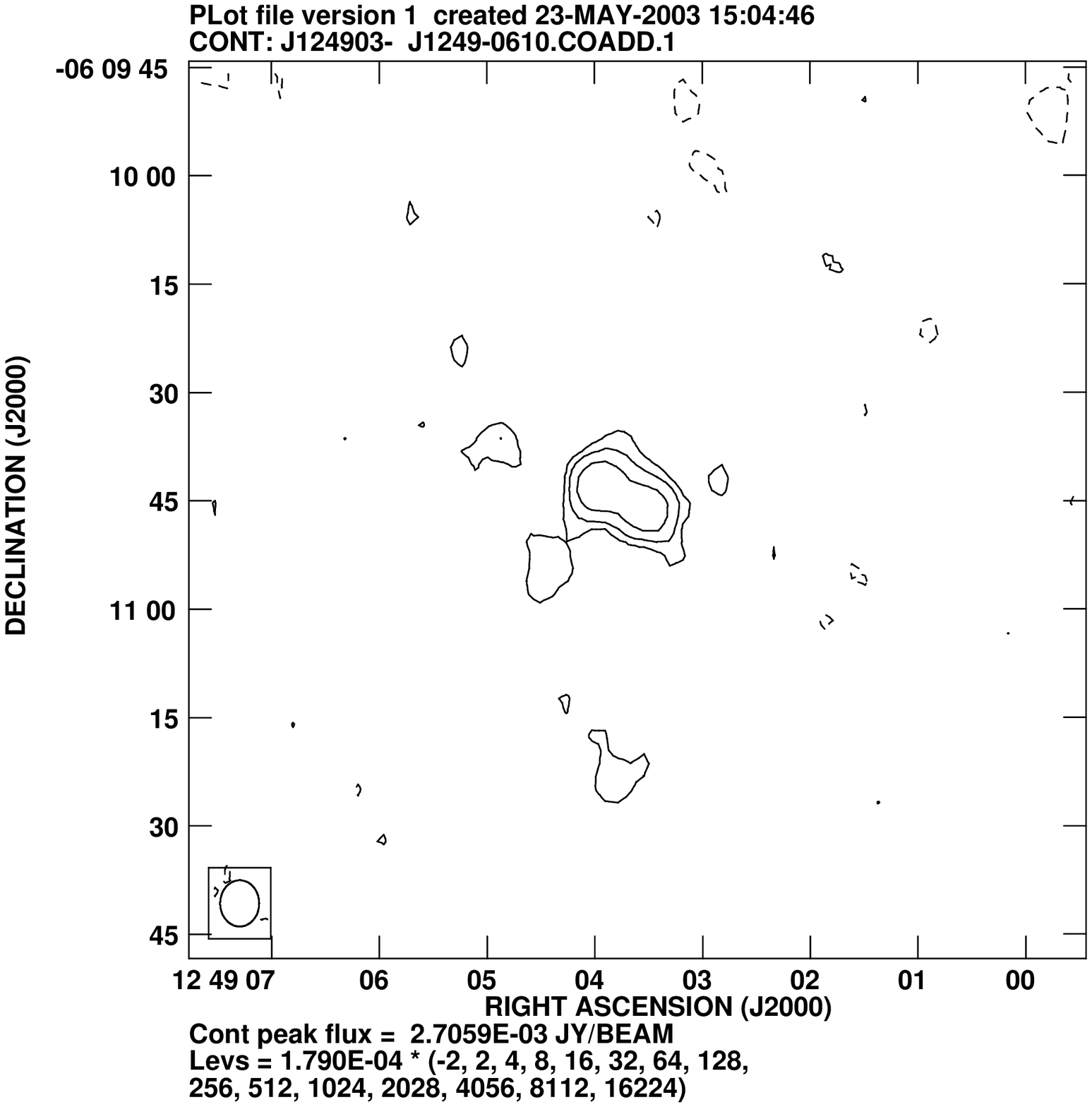}
\includegraphics[width=6cm]{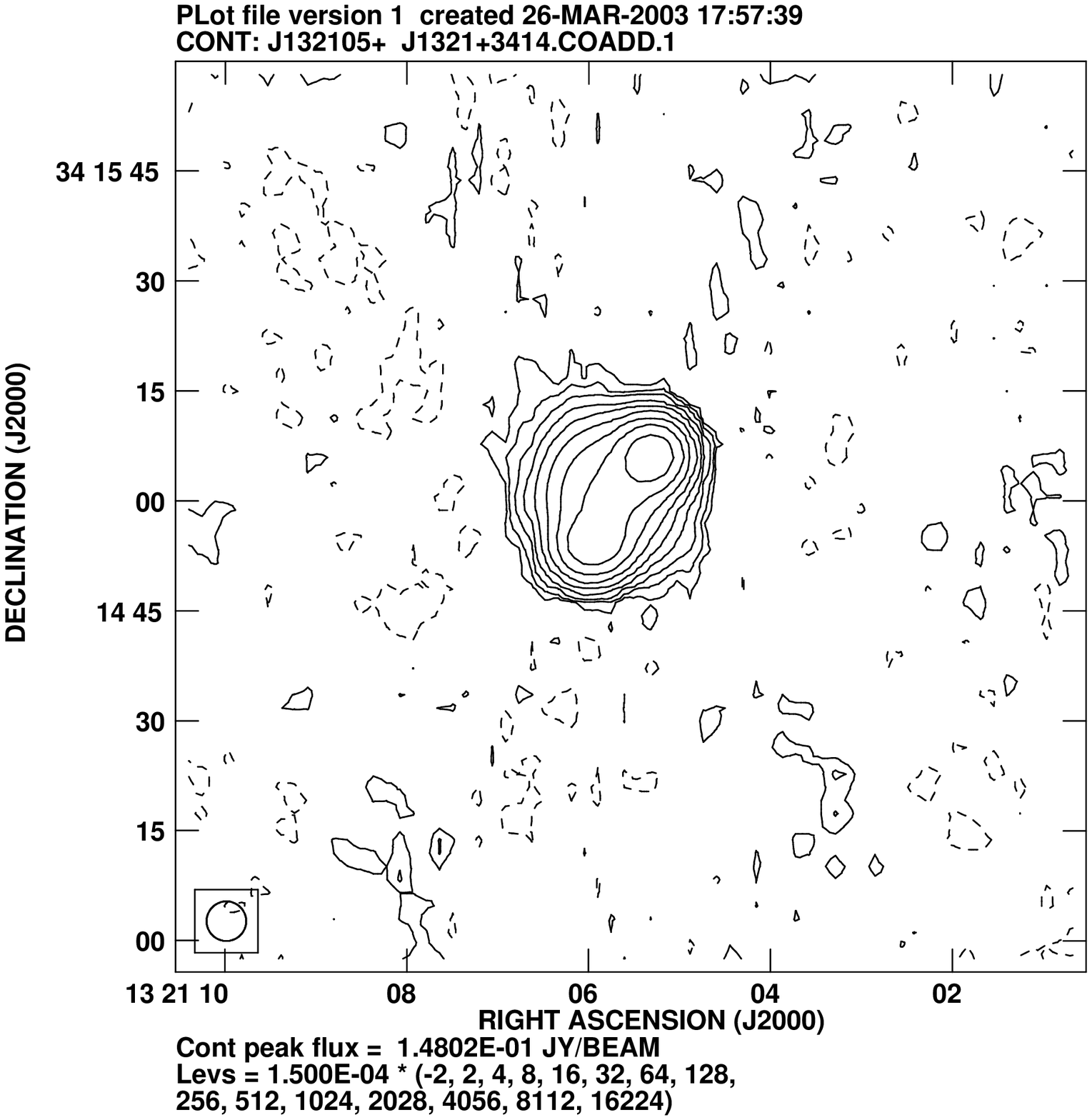}
\caption{Radio maps at 1.4 GHz of the 3 RL AGNs resolved in the FIRST
maps and showing double or triple morphology. From top to bottom: 
XBSJ111928.5+130250, XBSJ124903.6-061049 and XBSJ132105.5+341459}
\label{maps}
\end{figure}

The FIRST data are available for 12 sources out of 25 RL (8 of these have also fluxes at
other frequencies) and for the 3 sources RQ. From the FIRST survey 
we have taken a radio map at 1.4 GHz with a resolution
(beam size $\sim$5$\arcsec$) a  factor $\sim$10 better than that of the NVSS data. 
We have then calculated a core-dominance parameter based on the
peak and on the total flux derived from the FIRST catalogue. The parameter is defined as:

\begin{center}
core dominance =  {\Large $\frac{S_{core}}{S_{ext}}$}
\end{center}

where we have assumed the core flux density (S$_{core}$) equal to the
peak flux density of the brightest component, and the extended flux
density ($S_{ext}$) equal to the total flux density minus the peak
flux density.  The total flux density is the integrated flux density
taken from the FIRST catalogue.
For 3 sources (2 AGNs and 1 unidentified object) the
FIRST map shows a double or triple morphology and a core dominance $<$0.5$-$0.6
(see Figure~\ref{maps}).
These 3 objects are thus considered as ``non-blazars'', on the basis of the
radio morphology. We note that the remaining objects, that are compact at the
FIRST resolution, cannot be classified as ``blazar'' since the beam size is not necessarily
small enough to resolve the source, given the observed range of redshift.
For this reason, we use the morphological information only to classify
the 3 extended objects as ``non-blazars''.

\subsection{Summary of the radio classification}
In summary, based on the available radio or optical data, we have distinguished
the RL AGNs in the sample between ``blazars'' and ``non-blazars'' according to the
following criteria:

\begin{itemize}

\item All the objects optically classified as BL Lacs are considered blazars;

\item All the objects with a flat radio spectrum ($\alpha<$0.7) are classified
as blazars while all the sources with a steep radio spectrum ($\alpha>$0.7 or
a lower limit on $\alpha$ above 0.7) are classified as non-blazars. As discussed
above, there is one exception (XBSJ012654+191246), i.e. a powerful radiogalaxy that
has been classified as ``non-blazar'' on the basis of more detailed data
taken from the literature in spite of its relatively flat radio spectrum;

\item All the objects with a double or triple radio morphology (based on FIRST data)
are classified as non-blazars.

\end{itemize}

In total, we have divided the sample of radio-loud AGNs into 9 blazars,
10 non-blazars and 6 sources without a radio classification because of the lack of
specific data. This classification is reported in column 9 of Table~\ref{nvs}.
%

\section{The X-ray spectral analysis}

Thanks to the availability of good XMM-{\it Newton} data for the XBSS sample,
we can perform a reliable X-ray spectral analysis for each source in the sample.
The XBSS sample has been defined
using only the data from the MOS2 detector. In order to increase the
statistics for the X-ray spectra, the data from the MOS1 and the pn detectors, when available, 
are also considered and used for the analysis. 
Table~\ref{epic} shows the EPIC detectors used for the spectral analysis of each source along
with the values of Galactic column densities (\citealt{dickey90}) towards the selected XMM-{\it Newton} pointings.\\

    \begin{table*}
\begin{center}
\begin{small}
\caption{The XMM-{\it Newton} detectors used for the spectral analysis of the RL sample.}
\label{epic}
\begin{tabular}{c c l c c}
    \noalign{\smallskip}
         \hline
name&  obs ID&  Gal $N_H$ ($\times$10$^{20}$)& detector &  notes\\
(1) & (2) & (3) & (4) & (5) \\
    \noalign{\smallskip}
         \hline
 XBSJ000100.2$-$250501&  0125310101& 1.88& MOS2, MOS1, pn&  \\
 XBSJ003255.9+394619  &  0065770101& 6.57& MOS2, MOS1, pn&  \\
 XBSJ012505.4+014624 &  0109860101&3.10&MOS2, MOS1, pn&  \\
 XBSJ012654.3+191246 &  0112600601&4.80&MOS2, MOS1   & a  \\
 XBSJ013240.1$-$133307&  0084230301& 1.64& MOS2, MOS1, pn&    \\
 XBSJ013811.7$-$175416&  0111430101& 1.44& MOS2, MOS1    & a \\
 XBSJ021640.7$-$044404&  0112371701& 2.42& MOS2, MOS1, pn&    \\
 XBSJ033226.9$-$274107&  0108060501& 0.90& MOS2, MOS1, pn&    \\
 XBSJ052108.5$-$251913&  0085640101& 1.92& MOS2, MOS1, pn& b  \\
 XBSJ061342.7+710725&    0009220601& 8.38& MOS2, MOS1, pn&      \\
 XBSJ084026.2+650638&    0111400101& 4.29& MOS2, MOS1 &   \\
 XBSJ101511.8+520708&    0086750101& 0.76& MOS2, MOS1, pn& b \\
 XBSJ101706.5+520245&    0086750101& 0.76& MOS2, MOS1, pn&   \\
 XBSJ102016.1+082143&    0093640301& 2.99& MOS2, MOS1, pn&  \\
 XBSJ111928.5+130250&    0093641101& 2.43& MOS2, MOS1   & c\\
 XBSJ122658.1+333246&    0070340501& 1.39& MOS2, MOS1, pn& b \\
 XBSJ123538.6+621644&    0111550101& 1.49& MOS2, MOS1, pn&    \\
 XBSJ124903.6$-$061049&  0060370201& 2.13& MOS2, MOS1, pn&  \\
 XBSJ132105.5+341459&    0093640401& 1.00& MOS2, MOS1, pn& b\\
 XBSJ133026.6+241520&    0100240201& 1.16& MOS2, MOS1, pn&  \\
 XBSJ133232.6+111220&    0061940101& 1.93& MOS2, MOS1, pn&  \\
 XBSJ153452.3+013104&    0112190401& 4.89& MOS2, MOS1, pn&  \\
 XBSJ164237.9+030014&    0067340501& 5.47& MOS2, MOS1, pn& b\\
 XBSJ221951.6+120123&    0103861201& 5.41& MOS2, MOS1, pn&  \\
 XBSJ235036.9+362204&    0100241001& 8.09& MOS2, MOS1, pn& b\\
        \noalign{\smallskip}
            \hline
\end{tabular}
\end{small}
\end{center}
\footnotesize{{\bf column 1}: source name;
\\ {\bf column 2}: observation ID;
\\ {\bf column 3}: Galactic column density [cm$^{-2}$];
\\ {\bf column 4}: EPIC cameras used for the spectral analysis;
\\ {\bf column 5}: notes.\\ a= pn in small window.\\ b=source on a gap in the pn.
\\ c= source outside the pn.}
\end{table*}

The X-ray spectra usually cover the 0.2$-$10 keV energy range and have been
extracted using a circular region with a radius of
17.5$^{\prime\prime}$$-$25$^{\prime\prime}$. The background has been
extracted in nearby circular source free regions of an area a factor $\sim$4
larger than the one used to extract the source counts. In order to improve the statistics 
the MOS1 and MOS2 counts obtained with the same filters have been combined together.
The data of the MOS and of the pn detectors have
been binned in order to have at least 10$-$25 counts for each
channel, depending on the brightness of the source. For the
analysis of the data we used the XSPEC 11.2.0 software package. 
We have fitted simultaneously the MOS and pn data, 
leaving the relative normalization free to vary (range 0.9$-$1.2).  Errors are given at the
90\% confidence level for one interesting parameter ($\Delta\chi$$^{2}$=2.71). \\

\begin{table*}
\caption{Best-fit parameters for the X-ray spectral analysis.}
\label{x1}
\begin{tabular}{lllrrrlrr}
          \noalign{\smallskip}
          \hline
 name & id & model& $\Gamma$ & $N_H$ & kT & $\chi_{\nu}^{2}$/dof
& f$_{2-10keV}$ & Log(L$_{2-10keV}$) \\
(1) & (2) & (3) & (4) & (5) & (6) & (7) & (8) & (9) \\
      &    &         &          & ($\times$10$^{22}$) &          &   & ($\times$10$^{-13}$) & \\
            \noalign{\smallskip}
            \hline
 XBSJ000100.2$-$250501&  AGN1   &  PL   & 1.68$^{+0.2}_{-0.2}$    &  0.72$^{+0.23}_{-0.29}$
&   -                      &  1.19/40 &  1.39  & 44.67     \\
 XBSJ003255.9+394619&    AGN1   &  PL   & 2.11$^{+0.92}_{-0.41}$  &  $<$0.42
&   -              &  0.43/3  &  0.44  & 44.62     \\
 XBSJ012505.4+014624&    AGN1   &  PL   & 1.65$^{+0.06}_{-0.06}$  &  $<$0.47
&  -               &  0.86/101&  1.62  & 45.33    \\
 XBSJ012654.3+191246&    G      &  T    &   -          &  0.69$^{+8.24}_{-0.00}$
&  0.63$^{+0.02}_{-0.18}$   &  0.63/6  &  0.06  & 40.55   \\
 XBSJ013240.1$-$133307$^{(1)}$&  AGN2   &  PL   & 1.9*                &  2.54$^{+0.71}_{-0.56}$
&  -               &  1.11/19 &  1.71  & 44.41  \\
 XBSJ013811.7$-$175416&  BL?    &  PL   & 2.47$^{+0.25}_{-0.21}$  &  0.07 $^{+0.06}_{-0.06}$
&  -               &  1.40/31 &  0.36  & 43.68 \\
 XBSJ021640.7$-$044404$^{(2)}$&  AGN1   &  PL+BB & 2.28$^{+0.31}_{-0.26}$  &  0.56$^{+0.28}_{-0.56}$
&  0.15$^{+0.00}_{-0.13}$              &  1.08/93 &  1.14  & 44.73  \\
 XBSJ033226.9$-$274107&  AGN1   &  PL+BB & 2.04$^{+0.13}_{-0.15}$  &  $<$0.45
&  0.15$^{+0.06}_{-0.06}$          &  0.98/95 &  0.67  & 44.34  \\
 XBSJ052108.5$-$251913&  AGN1   &  PL+BB    & 1.72$^{+0.29}_{-0.19}$  &  0.10$^{+1.80}_{-0.10}$
& 0.23$^{+0.07}_{-0.04}$    & 0.90/61  &  2.46  & 45.26 \\
 XBSJ061342.7+710725&    BL     &  PL   & 2.67$^{+0.17}_{-0.15}$  &  0.14$^{+0.05}_{-0.32}$
&  -               & 0.79/69  &  1.87  & 43.75   \\
 XBSJ084026.2+650638&    AGN1   &  PL   & 1.71$^{+0.19}_{-0.18}$  &  0.29$^{+0.03}_{-0.2}$
&  -               &  1.08/34 &  1.15  &  44.89 \\
 XBSJ101511.8+520708&    AGN1   &  PL   & 2.06$^{+0.47}_{-0.31}$  &  $<$0.18
&  -               & 1.07/17  &  0.76  & 45.55    \\
 XBSJ101706.5+520245&    BL     &  PL   & 2.82$^{+0.83}_{-0.44}$  &  0.006$^{+0.008}_{-0.001}$
&  -               & 1.06/14  &  0.32  & 43.53    \\
 XBSJ102016.1+082143&    -       &  PL   & 1.97$^{+0.47}_{-0.23}$  &  $<$6.89E-2
&  -               & 0.40/17  &  0.67  & -  \\
 XBSJ111928.5+130250&    AGN1   &  PL   & 1.90$^{+0.33}_{-0.17}$  &  $<$0.62
&  -               & 0.95/6   &  1.11  & 45.74   \\
 XBSJ122658.1+333246&    CL    &  T & - &  $<$0.01
& 10.42$^{+1.19}_{-1.43}$                  & 0.93/205 & 3.06   & 45.14   \\
 XBSJ123538.6+621644&    AGN1   &  PL   & 1.96$^{+0.08}_{-0.05}$  &  $<$0.03
&  -               & 0.92/112 & 1.04   & 44.43    \\
 XBSJ124903.6$-$061049&  AGN1   &  PL   & 2.12$^{+0.18}_{-0.04}$  &  $<$0.03
&  -               & 1.20/70  & 0.87   & 44.35  \\
 XBSJ132105.5+341459&    -       &  PL   & 1.94$^{+0.15}_{-0.17}$  &  0.03$^{+0.03}_{-0.03}$
& -                &  1.18/53 &  1.64  &  - \\
 XBSJ133026.6+241520&    -       &  PL   & 2.74$^{+0.06}_{-0.04}$  &  $<$0.004
&  -               & 0.85/132 & 0.42   &  - \\
 XBSJ133232.6+111220&    -       &  PL   & 1.9*            &  $<$0.12
&  -               & 1.42/3   & 0.58   & -  \\
 XBSJ153452.3+013104&    AGN1   &  PL   & 1.75$^{+0.09}_{-0.04}$  &  $<$0.05
&  -               & 0.98/112 &  7.42  &  45.95  \\
 XBSJ164237.9+030014&    -       &  PL   & 1.62$^{+0.51}_{-0.14}$  &  $<$0.08
&  -               & 1.37/7   & 0.90   &  -  \\
 XBSJ221951.6+120123&    AGN2   &  PL   & 1.39$^{+0.36}_{-0.02}$  &  $<$0.35
&  -               & 1.26/12  & 1.26   &  44.10   \\
 XBSJ235036.9+362204&    BL     &  PL   & 1.90$^{+0.34}_{-0.22}$  &  0.03$^{+0.17}_{-0.03}$
& -                & 1.08/13  & 2.06   &  43.87   \\

           \noalign{\smallskip}
            \hline
\end{tabular}
\newline
\footnotesize{
{\bf column 1}: source name;

$^{(1)}$: The best fit values for this source are taken from \citet{caccianiga04};

$^{(2)}$: This source belongs also to the HBS28 sample discussed in \citet{caccianiga04}. In the X-ray analysis
presented here, however, we have added a soft excess component and the best fit parameters (the $N_H$ in particular) 
are newly calculated;
\\{\bf column 2}: optical classification;
\\{\bf column 3}: best-fit model (PL=absorbed power law; T=thermal component (mekal);
PL+BB=absorbed power law plus soft excess (modelled with a black-body spectrum);
\\{\bf column 4}: best-fit photon index of the power-law component (* in this case the value of the 
photon index has been fixed to 1.9);
\\{\bf column 5}: best fit intrinsic hydrogen column density [cm$^{-2}$];
\\{\bf column 6}: temperature of the thermal component or of the soft excess [keV];
\\{\bf column 7}: reduced Chi squared and degrees of freedom of the best fit;
\\{\bf column 8}: observed 2$-$10 keV flux (corrected for the Galactic absorption) [erg
cm$^{-2}$s$^{-1}$];
\\{\bf column 9}: intrinsic luminosity in the 2$-$10 keV energy range [erg s$^{-1}$].}
\end{table*}
\normalsize

We started the analysis of all the spectra by fitting a simple rest frame absorbed power law model,
including the Galactic absorption along the line of sight (derived from \citealt{dickey90}).

For the 2 sources for which the number of counts is
not high enough to accurately determine both the $\Gamma$ and the level of absorption,
we have fixed the value of $\Gamma$ to 1.9
in order to better constrain the correct value of
$N_H$. 
The value of $\Gamma$=1.9 corresponds to the average value found for the type 1 AGNs
in a complete and representative subsample of the ``hard'' XBSS (\citealt{caccianiga04}). 
As described in the next section, this value turned out to be slightly lower than the
average value derived from the analysis of the RL AGN sample (2.0$\pm$0.4) but
fully consistent with it within the errors.
These 2 sources are marked with an asterisk in column~4 of Table~4.

In total, for 20 sources the absorbed power law model describes
correctly the data. In three cases, instead, a soft-excess component,
modelled with a black-body spectrum with temperature ranging from 0.15 to 0.23 keV
is required in addition to the power-law component.
For the remaining 2 sources (XBSJ012654.3+191246 and XBSJ122658.1+333246)
the best fit is obtained with a single thermal spectrum. More details on these 2 objects
are given below.

{\bf XBSJ012654.3+191246}. For this source, optically classified as
``normal galaxy'' (see Figure~\ref{gal}), the simple absorbed
power law model gives a fit with strong residuals and a best-fit
slope too steep ($\Gamma$=4.6). If $\Gamma$ is frozen to
the mean value observed in AGNs ($\Gamma$=1.9) then the fit is
unacceptable ($\chi$$^{2}$/$\nu$=3.62, with 7 d.o.f.).
A model containing only a
thermal component (with kT=0.69 keV) gives a better fit.
In particular, we have used the spectrum from hot diffuse gas based
on the model ``mekal''. 
As discussed in Section~4.1, this is a known radiogalaxy.
The X-ray luminosity derived from this source (3.5$\times$10$^{40}$
erg s$^{-1}$) is within the typical range of X-ray luminosities observed
in low-power radiogalaxies (FR~I, e.g. \citealt{fabbiano84} and \citealt{fabbiano89}) .
\newline
{\bf XBSJ122658.1+333246}.
This source is a high-redshift (z=0.89) cluster of galaxies identified
in the WARPS survey (\citealt{ebeling01}). The X-ray spectral analysis of this object shows
that the X-ray spectrum is well described by a thermal component (mekal). We obtain a
temperature of kT=10.42$^{+1.19}_{-1.43}$ keV, which is in good agreement with that
obtained by \citet{cagnoni01} (kT=10.47$^{+4}_{-3}$ keV) with Chandra data and with that obtained from
the Sunyaev-Zeldovich measurement by \citet{joy01} (kT=10.00$^{+2.0}_{-1.5}$ keV). 
This is one of the highest temperatures ever measured in a cluster of galaxies. 
In conclusion, in this object the X-ray emission is associated to the hot intra-cluster medium 
while the radio and the optical emissions are relative to a radiogalaxy that belongs to the cluster.

The results of the
X-ray spectral analysis of the radio loud sources are reported in
Table~\ref{x1} and the X-ray spectra are reported in appendix. 
For one source (XBSJ013240.1$-$133307), which is optically classified
as type 2 AGN, the derived hydrogen column density is
significantly larger than the Galactic column density thus
confirming the presence of absorption also in the X-ray band. 
Given its X-ray luminosity 
(1.5$\times$10$^{44}$ erg s$^{-1}$)  and radio power 
(6$\times$10$^{24}$ W Hz$^{-1}$), this source can be classified as radio-loud 
type~2 QSO (see the discussion  in \citealt{dellaceca03} on a similar object 
discovered with ASCA). 
We note that for the other type~2 AGN present in the sample 
(XBSJ221951.6+120123) the X-ray analysis does not reveal a significant absorption
($N_H\leq$3.5$\times$10$^{21}$ cm$^{-2}$). 
However, the very flat best-fit photon index  ($\Gamma$=1.39) could be suggestive of the
presence of a very large absorption plus a flat reprocessed component. 
Therefore in this source the presence of a large 
absorption cannot be completely ruled out. Given the large X-ray
luminosity, this source is another type~2 radio-loud QSO candidate.

\begin{figure}
\centering
\includegraphics[width=9cm]{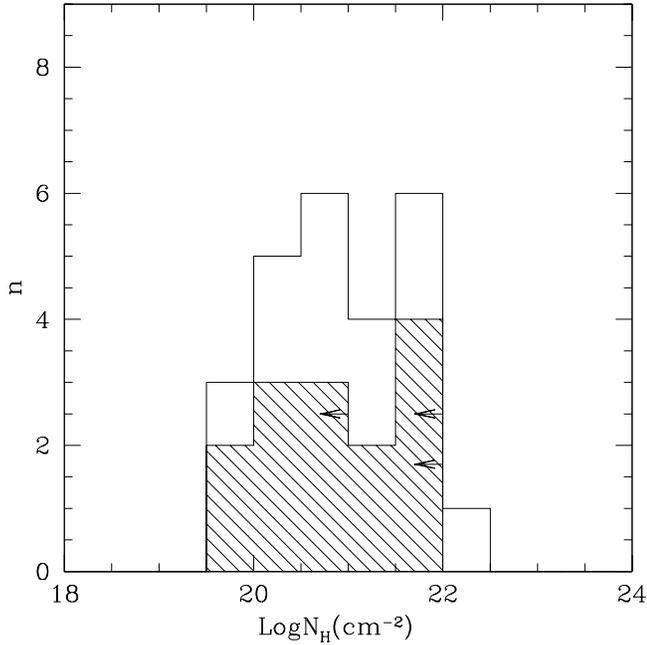}
\caption{The distribution of the intrinsic $N_H$ derived from the spectral fit.
The shaded histogram shows the upper limits on $N_H$ while the empty
histogram represents the detections.}
\label{nh}
\end{figure}

\begin{figure}
\centering
\includegraphics[width=9cm]{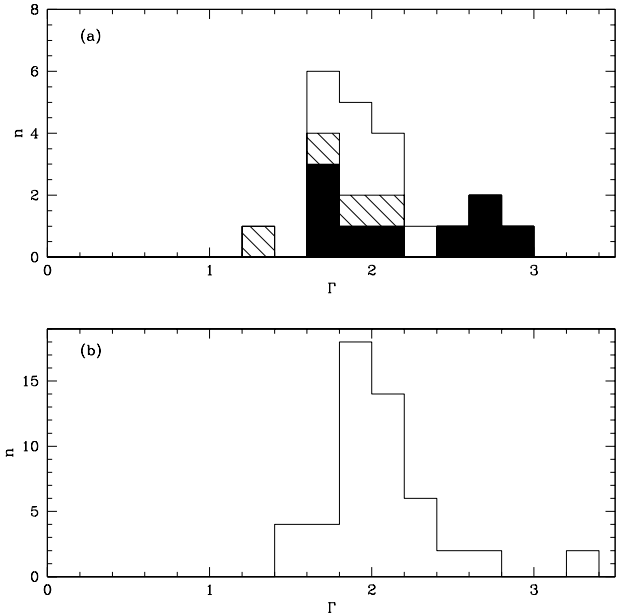}
\caption{The X-ray photon index distribution obtained from the spectral fitting for the
RL sources (a) and for the comparison sample (b). In Figure (a) the black histogram
represents the blazar, the empty histogram represents the non-blazar sources and the
shaded histogram shows the sources with no data to allow a radio classification.
In these histograms the sources for which the value of $\Gamma$ has been fixed to 1.9
(see text for details) and those fitted only by a thermal emission are not included.}
\label{gamma}
\end{figure}

The distribution of the
intrinsic $N_H$ derived from the spectral fit is shown in Figure~\ref{nh} 
while the distribution of the photon indices $\Gamma$ is shown in Figure~\ref{gamma}a.
The latter histogram does not include the  objects for which $\Gamma$ has been fixed
to 1.9. The mean value is 2.0 with a standard deviation of 0.4.

%

%

\section{Comparison between RL and RQ AGNs}

   \subsection{The comparison sample}

The 25 X-ray sources studied in this work are radio loud AGNs. In order to investigate the X-ray
spectral differences
between RL and RQ AGNs, we have defined a comparison sample.
This sample is selected from the
same XBSS catalogue and it consists of X-ray extragalactic 
sources which are not detected
in the NVSS survey, thus having a flux density below
2.5 mJy/beam.

In order to have a high number of sources spectroscopically
identified, we have considered  only the area of the sky
with declination $>$$-$30$^{\circ}$
and with right ascension 20$^{h}$$<\alpha<$2$^{h}$.
 In this area there are
67 sources out of which 65 have a spectroscopic classification.
For the comparison we use only the extragalactic subsample composed of 51 objects
plus the 2 unidentified sources. On the basis of their hardness-ratios and 
X-ray-to-optical flux ratios, in fact, these 2 sources are very likely to be
extragalactic objects (see \citealt{dellaceca04}). In conclusion, the comparison
sample contains 53 XBSS sources with a radio flux density at 1.4~GHz below 2.5 mJy.

The classification  breakdown of this comparison sample is presented in Table~\ref{confr}.
Apart from one cluster of galaxies and 2 ``normal'' galaxies, the remaining 48
objects are optically classified as AGNs. From the X-ray
spectral analysis we have direct evidence of the presence of an AGN
also in the two objects optically classified as ``normal'' galaxies.
The comparison sample is thus mainly composed by AGNs.

For a reliable analysis of the differences between RQ and RL AGNs 
it is important that the comparison sample is representative of the
RQ population. Given the lack of measured radio fluxes, for the sources in the comparison
sample it is only possible to compute an upper limit on the radio-loudness
parameter (see Section~2.1). In about 30\% of the sources the 
computed upper limit is below the dividing line between RQ and RL AGNs so that
they can be classified as RQ sources. For the remaining objects a firm
classification as RQ objects is not possible with the present data.
However, the analysis described in Section 2 of the distribution of the
radio-loudness parameter (and the upper limits) indicates that
among the sources not detected in the NVSS we expect a small
contamination of RL AGNs (about 2 objects in the comparison sample
defined here). Hence, even if these few RL AGNs cannot be identified 
without a deeper radio follow-up, we can consider
the whole comparison sample as statistically representative of the RQ AGN 
population. \newline

The X-ray spectra of the 53 objects contained in the comparison
sample have been extracted and analysed with the same procedures and
criteria described previously for the sample of RL AGNs.

\begin{figure}
\centering
\includegraphics[width=8cm]{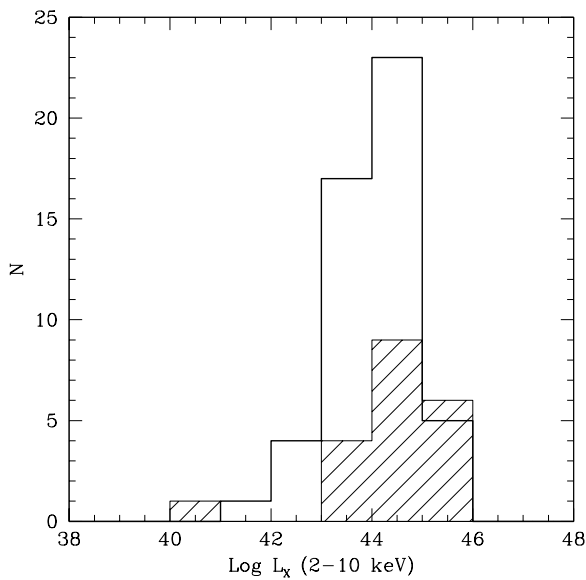}
\caption{Distribution of the de-absorbed 2-10 keV luminosity of 
the sample of RL AGNs (shaded histogram) and the
comparison sample (thick line). Only the sources in the 2 samples with a 
measured redshift are considered.}
\label{lx}
\end{figure}

In Figure~\ref{lx} we have reported the distribution of the de-absorbed X-ray 
luminosities (2$-$10 keV) of the two samples as resulting from the
X-ray spectral analysis. Although the RL AGNs have on average  
slightly higher luminosities (mean $L_X$=10$^{44.4}$ erg s$^{-1}$) than RQ 
AGNs (mean $L_X$=10$^{44.0}$ erg s$^{-1}$), the two distributions cover a very
similar range of values and the KS probability ($P_{KS}$=6\%) is only marginally 
indicative that the two distributions are not derived from the same parent 
distribution. For this reason, we consider the comparison sample as
a good choice to match the X-ray properties of RL and RQ AGNs since they 
both contain objects within a similar range of luminosities.

\subsection{Differences in the optical composition}
In Table~\ref{confr} the optical compositions of the radio-loud and of the
comparison samples are summarized and compared. The main difference between the
two samples rests in the percentage of BL Lac objects which is low (2\%) in
the comparison sample and much higher (20\%) in the sample of RL AGNs. 
We recall here that the only BL Lac
found in the comparison sample is not necessarily a RQ object since, as stated above, 
we expect a small percentage ($\sim$2 objects) of RL AGNs also in this sample. 
Thus, the percentage of BL Lacs in the 2 samples could be  even more different 
than what is reported in Table~\ref{confr}. 
This result confirms very clearly that BL Lac objects (at least, the X-ray selected ones)
are radio-loud AGNs, as firstly pointed out by \citet{stocke90}. Hence, if radio-quiet
BL Lacs exist, as recently suggested by \citet{londish04}, these objects must be
also ``X-ray-quiet'', i.e. they must have a very anomalous Spectral Energy Distribution (SED).

If we exclude the BL Lac objects from the 2 samples and re-normalize the percentages 
of the different classes of sources, the relative numbers are very similar in the two samples
(the number of clusters and normal galaxies is too small to have any statistical significance).
We conclude that the requirement of a radio emission does not change significantly the
composition of the sample, apart from the BL Lac objects that are greatly
favoured by the radio constraint. Indeed, the combination of X-ray and radio data
has been widely used in the last years to select large samples of BL Lacs (\citealt{maccacaro98}; 
\citealt{laurent-muehleisen98}; \citealt{perlman98}; \citealt{caccianiga99}; 
\citealt{giommi99}).

\subsection{Comparing the intrinsic X-ray spectral index}
We have then compared the X-ray photon indices of the power law component
in the two samples. The comparison of the spectral index 
gives in fact important pieces of information on the
similarities/differences of the primary source of the X-ray
emission. 

In the following analysis we have not included the sources for which the
value of $\Gamma$ has been fixed to 1.9 and those fitted only
by a thermal spectrum. 
Figure~\ref{gamma} shows the distribution of the best-fit
spectral indices. The mean value of $\Gamma$
is very similar in the two samples ($\Gamma_{RL}$=2.0 and $\Gamma_{RQ}=$2.1
respectively).

\begin{figure}
\centering
\includegraphics[width=6cm]{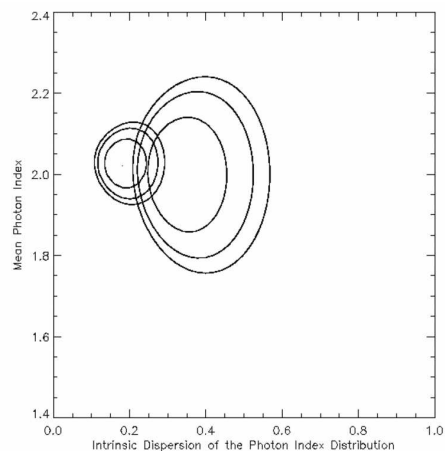}
\caption{Confidence contours (68\%, 90\%, and 95\%) for the joint
determination
of the $<\Gamma>$ and the intrinsic dispersion $\sigma$ of the photon
index distribution for the RL AGNs (curves on the right) and RQ AGNs (curves on the left).
In the analysis we have not included the sources with a fixed value of $\Gamma$ 
and those fitted only by a thermal emission.}
\label{contorni}
\end{figure}

We have then computed the intrinsic dispersion of the values of
$\Gamma$ following the method described in \citet{maccacaro88} which takes into
account the errors on the
values of $\Gamma$ (under the assumption of a Gaussian distribution).
In Figure~\ref{contorni} the confidence contours
(68\%, 90\%, and 95\%) for the joint determination of $<\Gamma>$
and of the intrinsic dispersion $\sigma$ are presented.
It is evident that, while the mean values of $\Gamma$ in the two samples
are very similar, the intrinsic spreads are
significantly (95\%) different ($\sigma_{RL}$=0.34 vs $\sigma_{RQ}$=0.18).

\begin{figure}
\centering
\includegraphics[width=6cm]{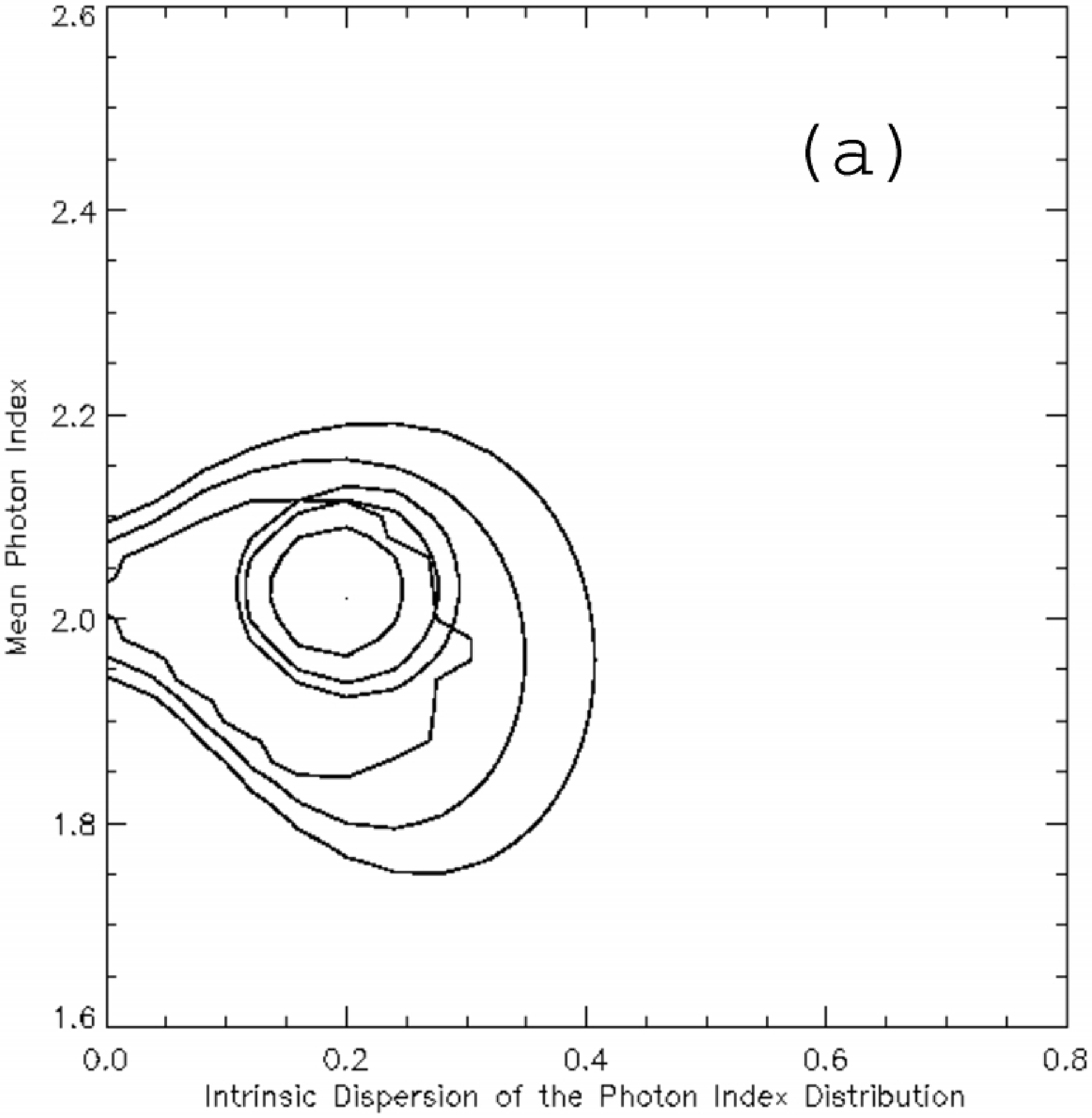}
\includegraphics[width=6cm]{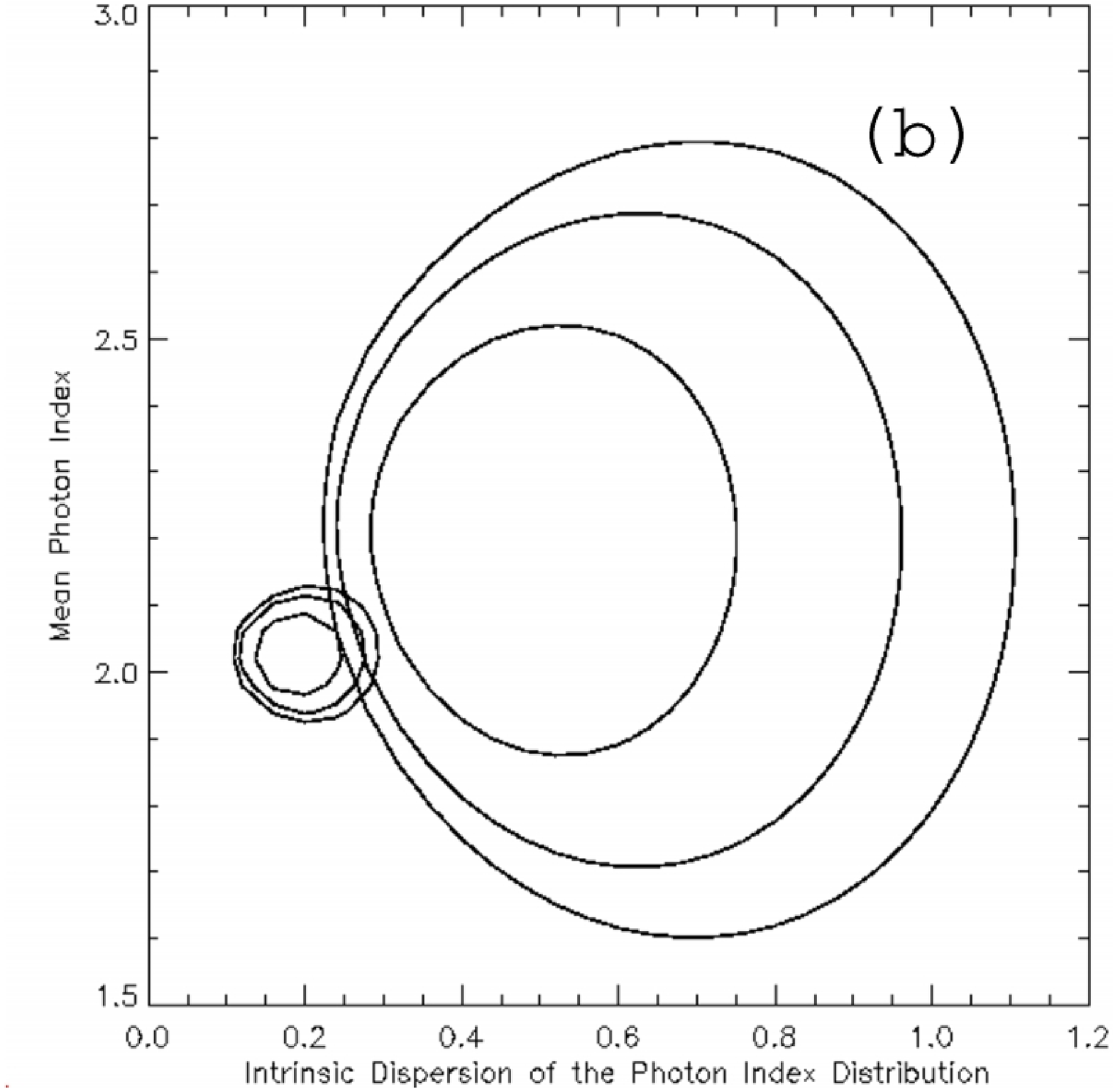}
\caption{The same as in Figure~\ref{contorni} but keeping separate the
non-blazars (panel a) and the blazars (panel b).}
\label{nonbl}
\end{figure}

We have analysed the possibility that the observed broader distribution
is due to the presence, in the RL AGN sample, of the
blazars. In fact, observing
the distribution presented in Figure~\ref{gamma} (a),
BL Lacs and the FSRQs are populating respectively the steep
and flat tails of the distribution ($<\Gamma>_{BL Lac}$=2.5 and
$<\Gamma>_{FSRQ}$=2.0).
We have thus  re-analysed the two samples considering separately
the non-blazars and the blazars (see Figure~\ref{nonbl}).
Although the number of objects is small, it is clear that
in the first case the two distributions are fully consistent,
both in terms of $<\Gamma>$ and in terms of intrinsic
spread ($\sigma_{RL}$=0.12 and $\sigma_{RQ}$=0.18) while, in the
case of blazars, the intrinsic spread is significantly larger ($\sigma_{blazar}$=0.44) 
than the one observed in RQ AGNs.

Overall, the results discussed here show that the distribution of $\Gamma$ of the
non-blazars is very similar, in terms of the mean value and the intrinsic spread,
to the one observed in RQ AGNs.
The class of blazars, instead, presents a much broader distribution of
values of $\Gamma$ (although the mean value is consistent with that observed
in the comparison sample) with the FSRQ and the BL Lac objects
populating respectively the flat and the steep tails of the distribution.
Therefore, the presence of a significant number of blazars in the RL sample broadens
the observed distribution of $\Gamma$ but it does not change the mean value.
In the next section we further investigate the origin of the X-ray emission of the blazars
present in the sample of RL AGNs.

\section{The X-ray spectra of Blazars}
According to the beaming model (e.g. \citealt{urry95}
and references therein)
the class of blazars is represented by RL AGNs observed very close to the
direction of the relativistic jets. The non-thermal emission produced within the
jet is thus relativistically amplified towards the observer becoming dominant
in most of the observing frequencies. The SED
of blazars is usually modelled with two large humps produced, respectively,
via synchrotron and Inverse Compton (IC) emission. The frequencies at which these 
two components peak can vary in a wide range of values leading to very different 
radio-to-optical and  optical-to-X-ray ratios: very large
values of the synchrotron peak frequency ($\nu>$10$^{15.5}$ Hz) typically produce objects
with very flat X-ray-to-radio spectral indices\footnote{
The two-point spectral index $\alpha_{RX}$ is defined in
the following way: $\alpha_{RX}=$ --Log (S$_R$/S$_X$)/Log ($\nu_R$/$\nu_X$),
where S$_R$, $\nu_R$ and S$_X$, $\nu_X$ are the monochromatic fluxes and frequencies
in the radio (5~GHz) and in the X-rays (2 keV),  respectively.}
($\alpha_{RX}<$0.7) while low values
of the peak frequencies ($\nu<$10$^{15.5}$ Hz) produce steep $\alpha_{RX}$ 
($\alpha_{RX}>$0.7).

The first systematic ROSAT observations of samples of blazars have revealed
a connection between the X-ray spectral index and the value of $\alpha_{RX}$:
objects with $\alpha_{RX}<$0.8 show steep values of $\Gamma$
($\sim$2.5) while blazars with $\alpha_{RX}>$0.8 have flat X-ray spectral indices
($<\Gamma>\sim$1.5-2.0; \citealt{padovani96}; \citealt{padovani97}; \citealt{urry96};
\citealt{lamer96}).
The connection between $\Gamma$ and $\alpha_{RX}$ (or between $\Gamma$ and
the synchrotron peak frequency if a SED is available) has been found also using the data
from BeppoSAX and ASCA (\citealt{wolter98}; \citealt{padovani01}; \citealt{donato01};
\citealt{beckmann02}; \citealt{padovani04}).

The usual interpretation of this trend is that the X-ray spectral index must
be affected by the energy at which the synchrotron component peaks: for large peak
frequencies, the X-ray spectrum includes the (steep) descending part of the synchrotron
emission while for low peak frequencies the X-ray spectrum includes the rising (and thus flatter)
IC component or a mixture of IC plus synchrotron emission.

In this framework it is interesting to see whether the 9 blazars selected in the
sample follow this trend: in this case, in fact, we would have a more
direct evidence that the X-ray emission in these sources is linked to the
radio emission and is thus produced within the jet.

\begin{figure}
\centering
\includegraphics[width=6cm]{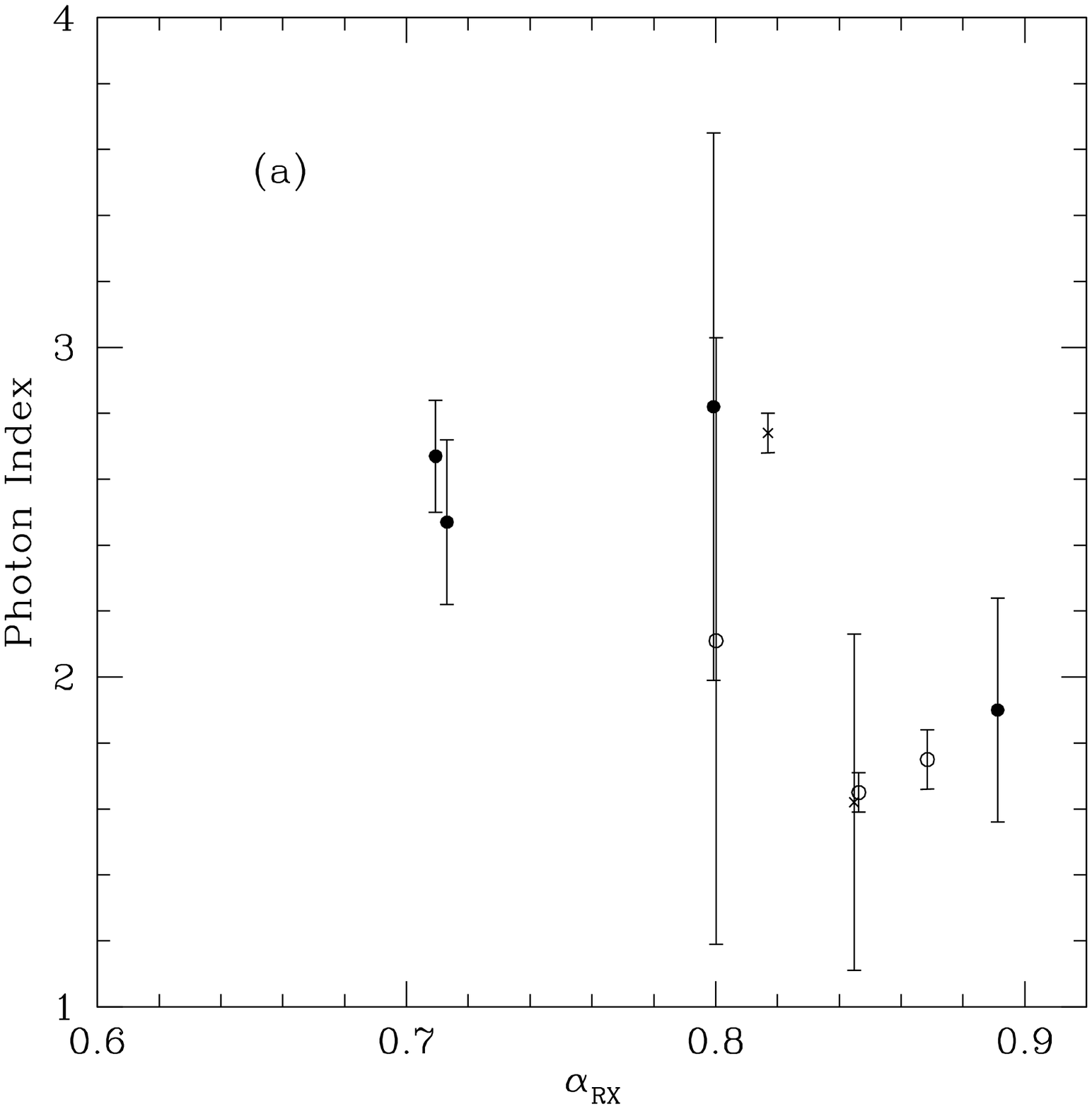}
\includegraphics[width=6cm]{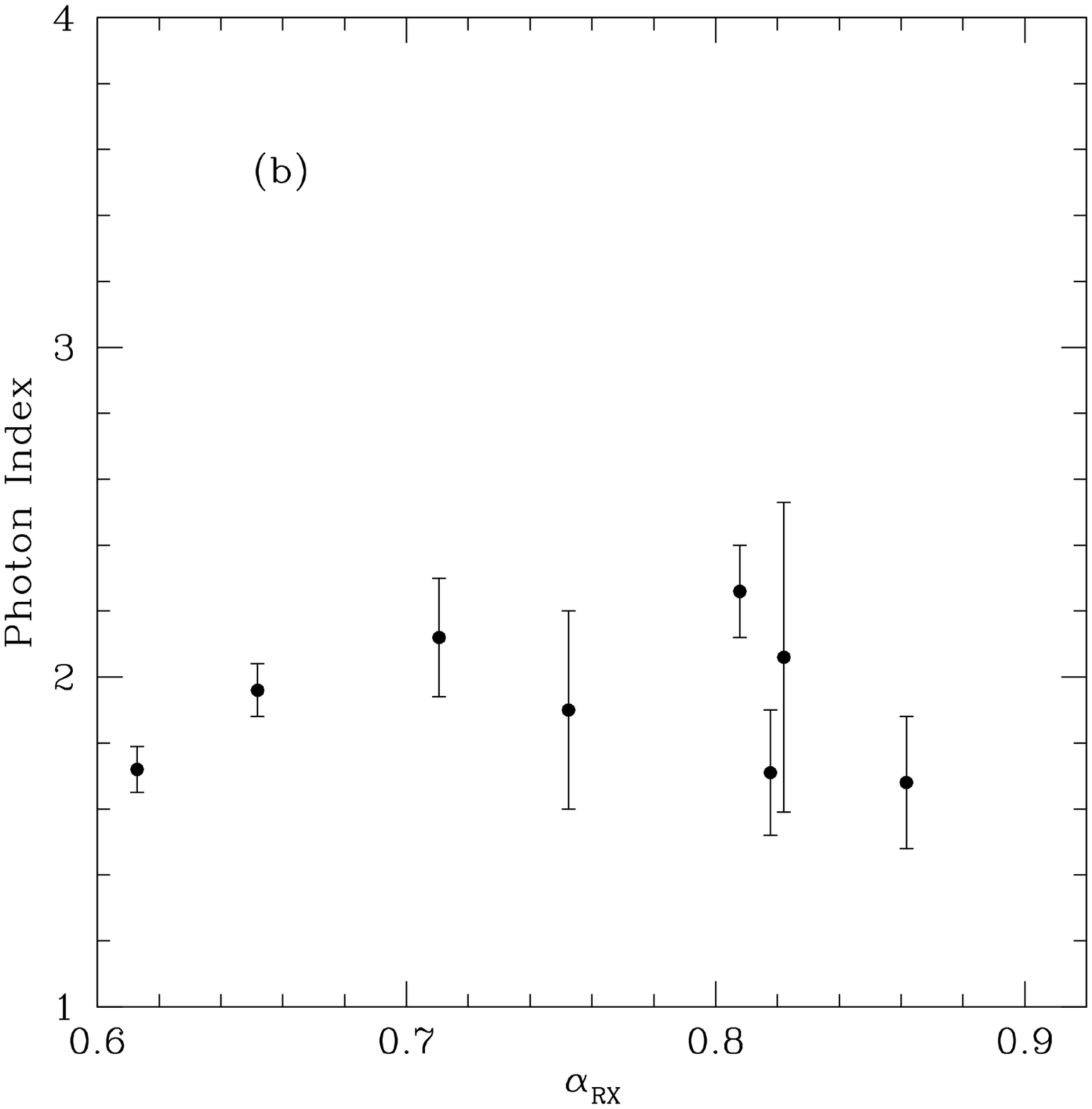}
\caption{The X-ray Photon index versus the radio-to-X-ray spectral index
($\alpha_{RX}$) for the sample of blazars (panel a) and the ``non-blazars''
(panel b).}
\label{gamma_arx}
\end{figure}

In Figure~\ref{gamma_arx}a we plot the best-fit photon indices of the 9 blazars
versus the $\alpha_{RX}$ values. Similarly to what is found in other blazar samples, the
objects characterized by low values of $\alpha_{RX}$ ($<$0.85) have also the
steepest X-ray spectra ($\Gamma\sim$2.5-2.8) while all the blazars with
large values of $\alpha_{RX}$ ($>$0.85) have a photon index below 2.

This trend is not found for the ``non-blazars''  (Figure~\ref{gamma_arx}b). 
In this case, the lack of trend is suggestive that the X-ray emission is
not directly related to the radio power, i.e. to the jet emission. This result
is in agreement with the idea that, in the ``non-blazars'', the relativistic beaming is not
playing an important role so that the jet contribution to the observed X-ray emission
is probably negligible and therefore the X-ray emission is more similar to that observed in 
RQ AGNs.

\section{Summary and conclusions}

We have presented the analysis of a sample of RL AGNs extracted from the
XBSS survey. The NVSS data  have been used to select
28 sources of the sample that are also radio emitting at 1.4~GHz with
a flux larger than 2.5~mJy. Out of these 28 radio/X-ray matches, 25 are classified 
as radio-loud AGNs 
on the basis of the radio-loudness parameter (R$>$10). The presence of an AGN
is detected in the optical and/or in the X-ray spectrum for 23 out of the 25 sources. 
In the remaining 2 sources, optically classified as ``galaxy'' and ``cluster'' respectively,
and for which the X-ray spectrum is modelled by a thermal emission, the presence of
an AGN (a radiogalaxy) is strongly suggested by the radio properties (radio-to-optical flux, 
the radio power or morphology). Therefore, the sample studied here is fully 
composed by AGN. 

Taking into account the flux limit of the NVSS (2.5 mJy/beam), we have estimated that the
RL AGNs represent 13\% of the X-ray selected AGNs at a flux limit of
$\sim$7$\times$10$^{-14}$ erg s$^{-1}$ cm$^{-2}$.
We have then selected from the same XBSS a ``comparison'' sample of 53 sources
not detected in the NVSS survey, i.e. with a radio flux density below 2.5 mJy/beam.
We have estimated that the large majority of the sources in this sample
is represented by RQ AGNs and that at most 2 objects could be RL objects.

The X-ray spectra of the sources in both samples have been extracted, studied and
compared.

The main results can be summarized as follows:

\begin{itemize}

\item The spectra of the majority of the sources in both samples ($\geq$90\%)
are well described by a single power-law. The mean spectral index is similar in
both samples ($\Gamma\sim$2).

\item  Although the mean value of $\Gamma$ is similar, the
intrinsic spread of the values of  $\Gamma$ (i.e. the one computed taking
into account the uncertainties) is significantly different in the two samples,
with the RL AGNs having a broader
distribution ($\sigma$=0.34) when compared to the RQ AGNs ($\sigma$=0.18).

\item When blazars (BL Lacs and FSRQs) and ``non-blazars'' (radiogalaxies and
SSRQs) in the RL AGN sample are analysed
separately, the larger spread of the spectral index is observed only
in the blazars sample while the ``non-blazars'' show a distribution fully
consistent with the one observed in the RQ AGNs.
In particular, BL Lacs and FSRQ are populating respectively the steep
and the flat tails of the  distribution of $\Gamma$.

\item The values of the X-ray spectral index of blazars are correlated with
the type of SED, i.e. with the radio-to-X-ray spectral index ($\alpha_{RX}$).
Such a correlation is not observed in the RL AGNs classified
as ``non-blazars''. This result suggests that the X-ray emission in
the blazar class is directly linked, as expected, to the jet emission while this is probably
not the case for the ``non-blazars'', i.e. those RL AGNs whose relativistic
jet is not oriented towards the observer. In these objects the relativistic beaming
is not important and the observed X-ray emission is likely to have the same origin as
in the RQ AGNs.

\end{itemize}

In conclusion, the analysis presented here has shown that the distribution of the
X-ray spectral indices observed in a sample of RL AGNs strongly depends on the
importance of relativistic beaming in the selected sources, i.e. on the
type of RL AGNs included in the sample under analysis. First, the inclusion
of a significant fraction of ``oriented'' RL AGNs, i.e. of blazars, significantly
broadens the distribution. Second, the type of blazars included in the analysis
changes the observed distribution of X-ray spectral indices: if the sample selection favours
objects with low values of $\alpha_{RX}$ (like in a typical X-ray selected survey),
the distribution will be more populated towards steep X-ray indices.
Samples that include more sources with large values of $\alpha_{RX}$
(like in a typical radio survey) are likely to include more blazars 
with a flat X-ray spectral index.

\begin{acknowledgements}
This research has made use of the NASA/IPAC Extragalactic Database
(NED) which is operated by the Jet Propulsion Laboratory,
California Institute of Technology, under contract with the
National Aeronautics and Space Administration.
EG, AC, TM, VB, RDC and PS acknowledge partial
financial support by the Italian Space Agency (ASI grants:
I/R/062/02 and I/R/071/02) and by the MIUR
(Cofin-03-02-23). PS and AC acknowledge financial support by the {\it
Istituto Nazionale di Astrofisica} (INAF).
Based on observations collected at the Italian
``Telescopio Nazionale Galileo" (TNG), at the German-Spanish
Astronomical Center, Calar Alto (operated jointly by Max-Planck
Institut  f\"{u}r Astronomie and Instututo de Astrofisica de
Andalucia, CSIC) and at the European Southern Observatory (ESO).
The TNG telescope is operated in the island of La
Palma by the Centro Galileo Galilei of the INAF in the Spanish Observatorio
del Roque
de los Muchachos of the Instituto de Astrof\'{\i}sica de Canarias.
We would like to thank C. Paizis, M.J.M. March\~a, A. Wolter and L. Ballo 
for useful suggestions and comments.
We finally thank the APM team for maintaining their facility.
\end{acknowledgements}

\newpage

\appendix
\section{The X-ray spectra}
\begin{figure*}
\centering
\includegraphics[width=17cm]{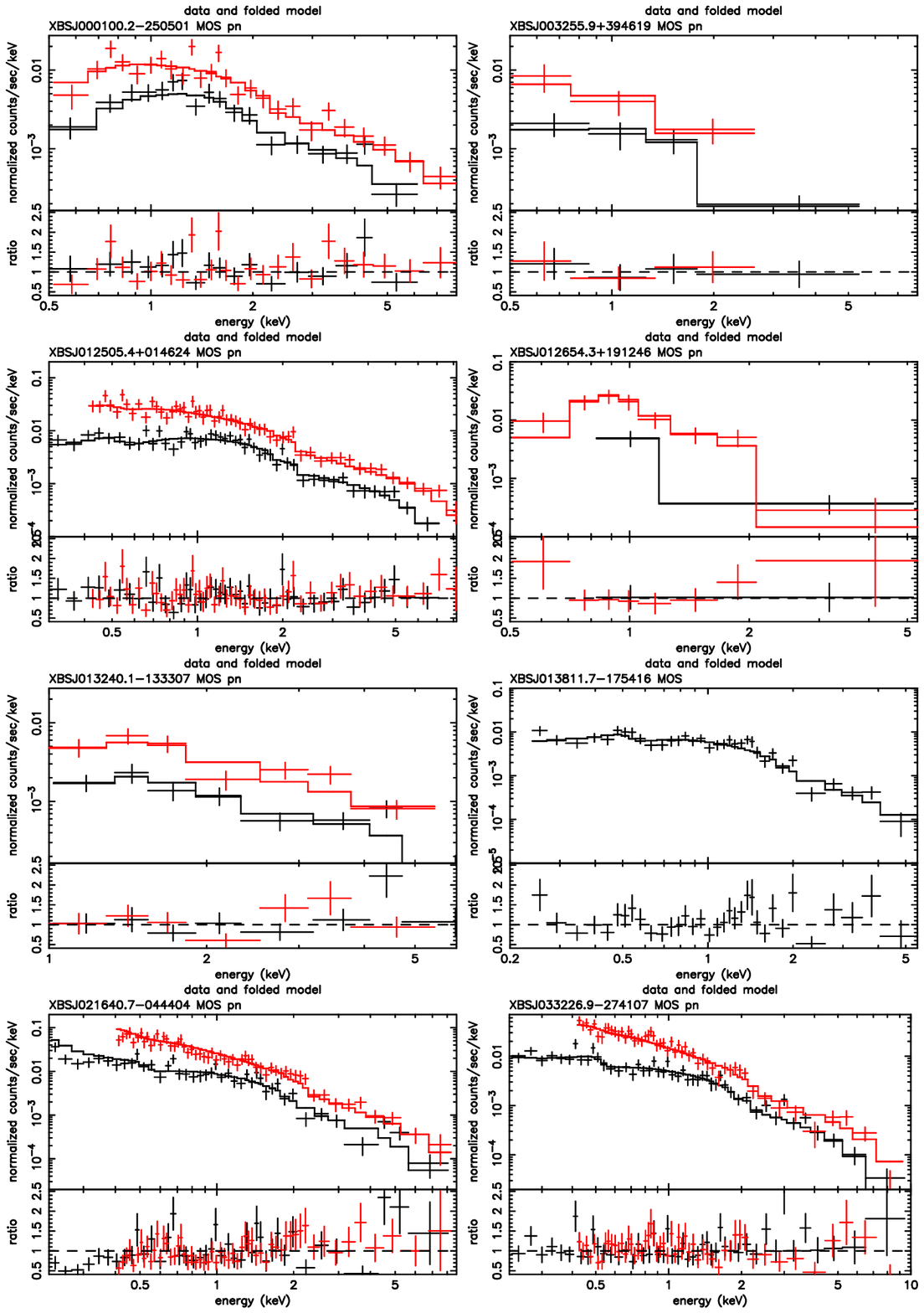}
\caption{The best-fit X-ray spectra of the 25 RL sources in the
 sample. Data (points) and folded model (continuous lines) are 
reported.
\label{xspec}}
\end{figure*}

\begin{figure*}[h]
\centering
\includegraphics[width=17cm]{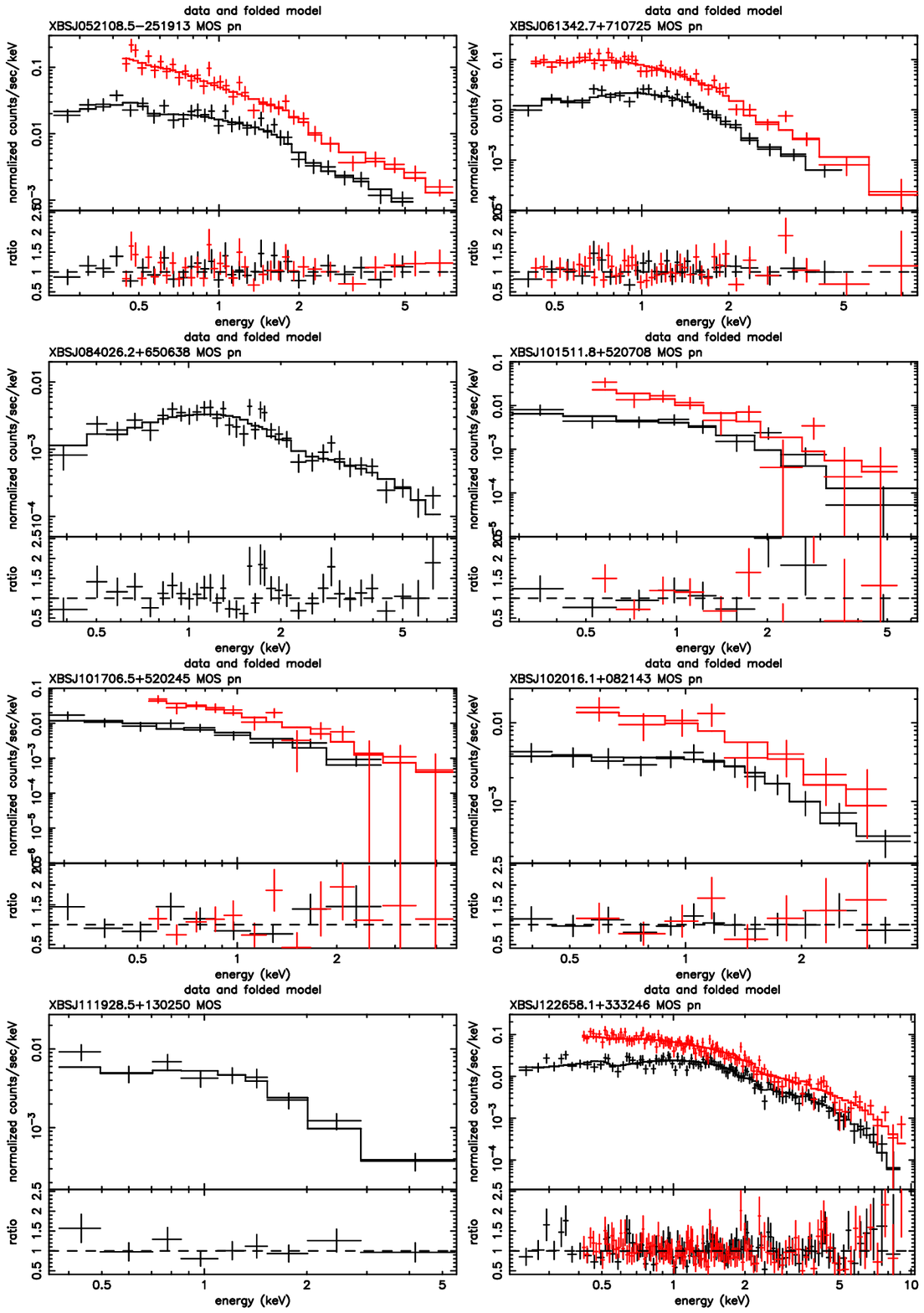}
\addtocounter{figure}{-1}
\caption{(continued)}
\end{figure*}

\begin{figure*}[h]
\centering
\includegraphics[width=17cm]{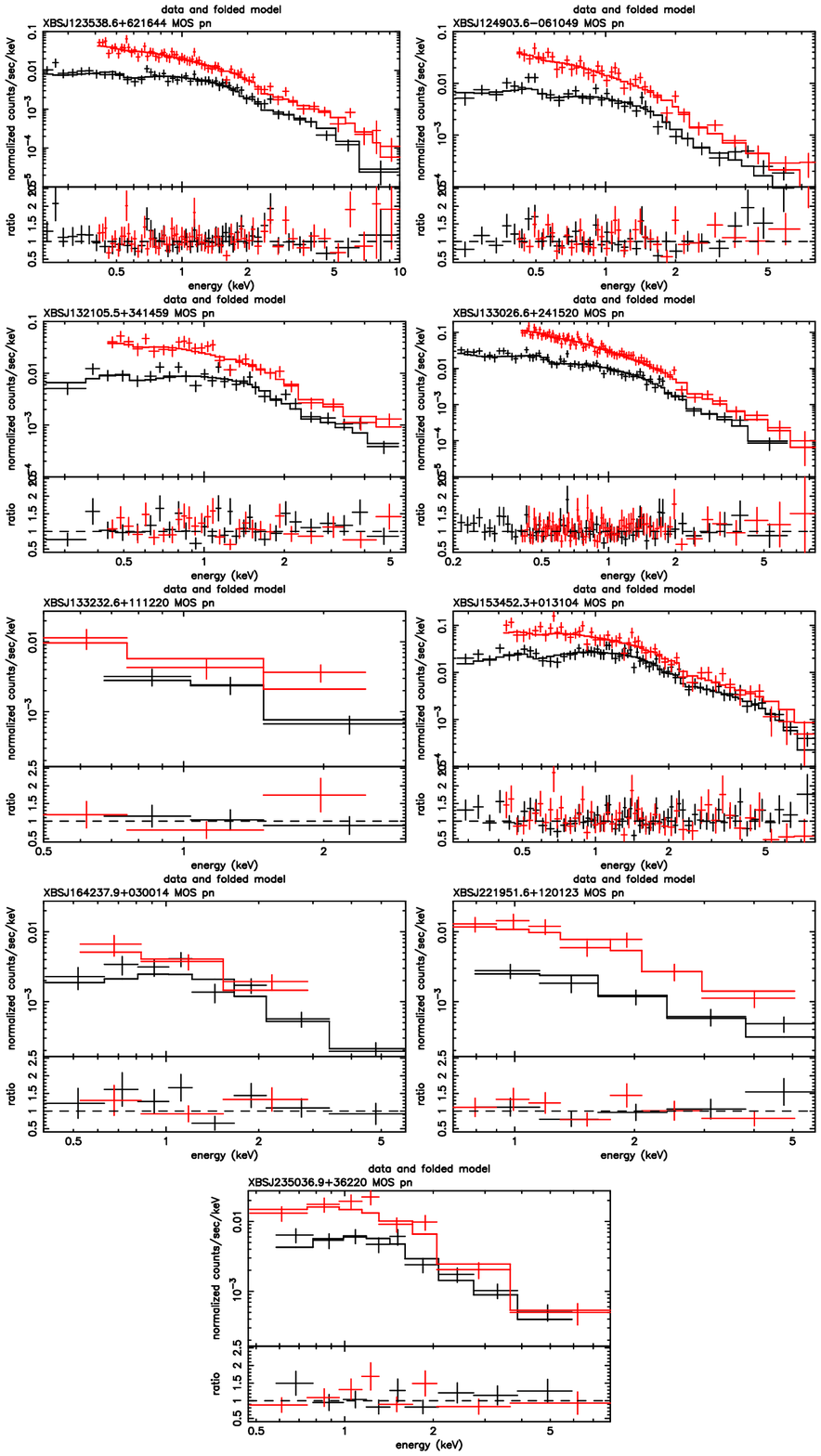}
\addtocounter{figure}{-1}
\caption{(continued)}
\end{figure*}



\begin{thebibliography}{99}
\bibitem[Avni et al. (1980)]{avni80}
Avny, Y., Soltan, H., Tananbaum, H., Zamorani, G., 1980, ApJ, 238, 800.
\bibitem[Ballantyne, Ross \& Fabian (2002)]{ballantyne02}
Ballantyne, D.R., Ross, R.R., Fabian, A.C., 2002, MNRAS, 336, 867.
\bibitem[Becker, White \& Helfand (1995)]{becker95}
Becker, R.H., White, R.L. \& Helfand, D.J., 1995, AJ, 450, 559.
\bibitem[Beckmann et al.(2002)]{beckmann02}
Beckmann, V., Wolter, A., Celotti, A., et al., 2002, A\&A, 384, 410
\bibitem[Brocksopp et al.(2004)]{brocksopp04}
Brocksopp, C., Puchnarewicz, E. M., Mason, K.O., C\'ordova, F.A., Priedhorsky, W.C., 2004, MNRAS, 349, 687.
\bibitem[Brunner et al.(1994)]{brunner94}
Brunner, H., Lamer, G., Worrall, D. M., Staubert, R., 1994, A\&A, 287, 436.
\bibitem[Caccianiga et al. (1999)]{caccianiga99}
Caccianiga, A.,  Maccacaro, T., Wolter, A., Della Ceca, R., Gioia, I.M., 1999, ApJ, 513, 51.
\bibitem[Caccianiga et al. (2001)]{caccianiga01} Caccianiga A., March\~a M.J.M., Ant\'on A., Mack
K.-H., Neeser M., 2001, MNRAS, 329, 877 
\bibitem[Caccianiga et al. (2004)]{caccianiga04}
Caccianiga, A., Severgnini, P., Braito, et al., 2004, A\&A, 416, 901.
\bibitem[Cagnoni et al. (2001)]{cagnoni01}
Cagnoni, I., Elvis, M., Kim, D.W., et al., 2001, ApJ, 560, 86.
\bibitem[Canizares \& White (1989)]{canizares89}
Canizares, C.R. \& White, J.L., 1989, ApJ, 339, 27.
\bibitem[Cavallotti et al. (2004)]{cavallotti04}
Cavallotti, F., Wolter, A., Stocke, J.T., Rector, T., 2004, A\&A, 419, 459.
\bibitem[Ciliegi et al. (2003)]{ciliegi03}
Ciliegi, P., Vignali, C., Fiore, F., La Franca, F., Perola, G.C., 2003, MNRAS, 342, 575.
\bibitem[Cirasuolo et al. (2003)]{cirasuolo03}
Cirasuolo, M., Celotti, A., Magliocchetti, M., Danese, L., 2003, MNRAS, 346, 447.
\bibitem[Condon et al. (1998)]{condon98}
Condon, J.J., Cotton, W.D., Greisen, E.W., et al., 1998, AJ, 115, 1693.
\bibitem[Della Ceca et al. (1994)]{dellaceca94}
Della Ceca, R., Zamorani, G., Maccacaro, et al., 1994, ApJ, 533, 544.
\bibitem[Della Ceca et al.(2003)]{dellaceca03}
Della Ceca, R., Braito, V., Beckmann, V., et al. 2003, A\&A, 406, 555 
\bibitem[Della Ceca et al.(2004)]{dellaceca04}
Della Ceca, R., Maccacaro, T., Caccianiga, A., et al., 2004, A\&A, in press
\bibitem[Dickey \& Lockman (1990)]{dickey90}
Dickey, J. \&  Lockman, F.J., 1990, ARA\&A, 28, 215.
\bibitem[Donato et al. (2001)]{donato01}
Donato, D., Ghisellini, G., Tagliaferri, G., Fossati, G., 2001, A\&A, 375, 739.
\bibitem[Douglas et al. (1996)]{douglas96}
Douglas, J.N., Bash, F.N., Boyzan, F.A., Torrence, G.W., Wolfe, C., 1996, AJ, 111, 1945.
\bibitem[Ebeling et al. (2001)]{ebeling01}
Ebeling, H., Jones, L.R., Fairley, B.W., et al., 2001, ApJ, 548, 23.
\bibitem[Eracleous, Sambruna \& Mushotzky 2000]{eracleous00}
Eracleous, M., Sambruna, R., Mushotzky, R. F.,2000, ApJ, 537, 654.
\bibitem[Fabbiano et al. (1984)]{fabbiano84}
Fabbiano, G., Trinchieri, G., Elvis, M., Miller, L., Longair, M., 1984, ApJ, 277, 115
\bibitem[Fabbiano (1989)]{fabbiano89}
Fabbiano, G., 1989, ARA\&A, 27, 87.
\bibitem[Ferrero \& Brinkmann (2003)]{ferrero03}
Ferrero, E., Brinkmann, W., 2003, A\&A, 402, 465.
\bibitem[Fiore et al. (2003)]{fiore03}
Fiore, F., Brusa, M., Cocchia, F. et al., 2003, A\&A, 409, 79
\bibitem[Franceschini, Vercellone \& Fabian (1988)]{franceschini98}
Franceschini, A., Vercellone, S., Fabian, A.C., 1998, MNRAS, 297, 817.
\bibitem[Gambill et al.(2003)]{gambill03}
Gambill, J. K., Sambruna, R. M., Chartas, G., et al., 2003, A\&A, 401, 505.
\bibitem[Giommi, Menna \& Padovani (1999)]{giommi99}
Giommi, P., Menna, M. T., Padovani, P., 1999, MNRAS, 310, 465
\bibitem[Gregory et al. (1996)]{gregory96}
Gregory, P.C., Scott, W.K., Douglas, K. and Condon, J.J., 1996, ApJS, 103, 427.
\bibitem[Kellermann et al. (1989)]{kellermann89}
Kellermann, K.I., Sramek, R., Schmidt, M., Shaffer, D.B., Green, R., 1989, AJ, 98, 1195.
\bibitem[Hasenkopf, Sambruna \& Eracleous (2002)]{hasenkopf02}
Hasenkopf, C. A., Sambruna, R. M., Eracleous, M., 2002, ApJ, 575, 127.
\bibitem[Ho et al. (1997)]{ho97}
Ho, L.C., Filippenko, A.V., Sargent, W.L.W., 1997, ApJS, 112, 315
\bibitem[Ho \& Peng (2001)]{ho01}
Ho, L.C., \& Peng, C.Y., 2001, ApJ, 555, 650
\bibitem[Ho (2002)]{ho02}
Ho, L.C., 2002, ApJ, 564, 120.
\bibitem[Ivezic et al. (2002)]{ivezic02}
Ivezic, Z.,  Menou, K.,  Knapp, G.R. et al., 2002, AJ, 124, 236.
\bibitem[Joy et al. (2001)]{joy01}
Joy, M., LaRoque, S., Grego, L., et al.,  2001, ApJ, 551, L1.
\bibitem[Lamer, Brunner \& Staubert (1996)]{lamer96}
Lamer, G., Brunner, H., Staubert, R., 1996, A\&A, 311, 384.
\bibitem[Landt et al. (2002)]{landt02}
Landt, H., Padovani, P., Giommi, P., 2002, MNRAS, 336, 945.
\bibitem[Landt et al. (2001)]{landt01}
Landt, H., Padovani, P., Perlman, E.S., et al., 2001, MNRAS, 323, 757
\bibitem[Laor (2000)]{laor00}
Laor, A., 2000, ApJ, 543, 111.
\bibitem[Laurent-Muehleisen et al. (1998)]{laurent-muehleisen98}
Laurent-Muehleisen, S.A., Kollgaard, R.I., Ciardullo, R., et al., 1998, ApJS 118, 127.
\bibitem[Londish et al. (2004)]{londish04}
Londish, D., Heidt, J., Boyle, B. J., Croom, S. M., Kedziora-Chudczer, L., 2004, MNRAS, in press
\bibitem[Maccacaro et al. (1988)]{maccacaro88}
Maccacaro, T., Gioia, I.M., Wolter, A., Zamorani, G., Stocke, J.T., 1988, ApJ, 326, 680.
\bibitem[Maccacaro et al. (1998)]{maccacaro98}
Maccacaro, T., Caccianiga, A., Della Ceca, A., Wolter, A., Gioia, I. M., 1998, AN, 319, 15
\bibitem[March\~a et al. (1996)]{marcha96} March\~a M.J.M., Browne I.W.A., Impey C.D., Smith P.S.,
1996, MNRAS, 281, 425
\bibitem[March\~a et al. (2001)]{marcha01} March\~a M.J.M., Caccianiga A., Browne I.W.A., Jackson
N., 2001, MNRAS, 326, 1455
\bibitem[Meier (1999)]{meier99}
Meier, D.L., 1999, ApJ, 522, 753.
\bibitem[Morris et al. (1991)]{morris91}
Morris, S.L., Stocke, J.T., Gioia, I.M., et al., 1991, ApJ, 380, 49.
\bibitem[Nagao, Murayama \& Taniguchi (2001)]{nagao01}
Nagao, T., Murayama, T., Taniguchi, Y., 2001, ApJ, 546, 744
\bibitem[Owen \& Ledlow (1997)]{owen97}
Owen, F.N., Ledlow, M.J., 1997, ApJS, 108, 41.
\bibitem[ Padovani \& Giommi (1996)]{padovani96}
Padovani, P., Giommi, P., 1996, MNRAS, 279, 526
\bibitem[ Padovani, Giommi \& Fiore (1997)]{padovani97}
Padovani, P., Giommi, P., Fiore, F., 1997, MNRAS, 284, 569.
\bibitem[Padovani et al. (1999)]{padovani99}
Padovani, P., Morganti, R., Siebert, J., Vagnetti, F. and Cimatti, A., 1999, MNRAS, 304, 829.
\bibitem[Padovani et al. (2001)]{padovani01}
Padovani, P., Costamante, L,. Giommi, P., et al., 2001, MNRAS, 328, 931.
\bibitem[Padovani et al. (2004)]{padovani04}
Padovani, P., Costamante, L., Giommi, P., et al., 2004, MNRAS, 347, 1282.
\bibitem[Page et al.(2004)]{page04}
Page, K. L., Turner, M. J. L., Done, C., et al., 2004, MNRAS, 349, 57
\bibitem[Perlman et al. (1998)]{perlman98}
Perlman, E.S., Padovani, P., Giommi, P., et al., 1998, AJ, 115, 1253.
\bibitem[Reeves et al. (1997)]{reeves97}
Reeves, J. N., Turner, M.J.L., Ohashi, T.,  Kii, T., 1997,  MNRAS, 292, 468.
\bibitem[Reeves \& Turner (2000)]{reeves00}
Reeves, J.N., Turner, M.J L, 2000, MNRAS, 316, 234.
\bibitem[Rengelink et al. (1997)]{rengelink97}
Rengelink, R.B., Tang, Y., De Bruyn, A.G., et al., 1997, A\&AS, 124, 259.
\bibitem[Severgnini et al. (2003)]{severgnini03}
Severgnini, P., Caccianiga, A., Braito, et al.\ 2003, A\&A, 406, 483
\bibitem[Sambruna, Eracleous \& Mushotzky (1999)]{sambruna-er-m99}
Sambruna, R.M., Eracleous, M., Mushotzky, R., 1999, ApJ, 526, 60
\bibitem[Schwartz (2004)]{schwartz04}
Schwartz, D.A., 2004, in proceedings of the JENAM-2003 Symposium,
"Radio Astronomy at 70: from Karl Jansky to microjansky," Budapest, Hungary, 27-30 August 2003, EDP Sciences, eds.
L. Gurvits, S. Frey, and S. Rawlings
\bibitem[Shastri et al. (1993)]{shastri93}
Shastri, P., Wilkes, B.J., Elvis, M., McDowell, J., 1993, ApJ, 410, 29.
\bibitem[Stocke et al. (1990)]{stocke90}
Stocke, J.T., Morris, S.L., Gioia, I., et al., 1990, ApJ, 348, 141
\bibitem[Urry \& Padovani (1995)]{urry95}
Urry, C.M. \& Padovani, P., 1995, PASP, 107, 803.
\bibitem[Urry et al. (1996)]{urry96}
Urry, C.M., Sambruna, R.M., Worrall, D.M., et al., 1996, ApJ, 463, 424.
\bibitem[Visvanathan \& Wills (1998)]{visvanathan98}
Visvanathan, N., Wills, B.J., 1998, AJ, 116, 2119.
\bibitem[White et al. (2000)]{white00} White, R.L., et al., 2000, ApJS, 126, 133
\bibitem[Wilkes \& Elvis (1987)]{wilkes87}
Wilkes, B.J., Elvis, M., 1987, ApJ, 323, 243.
\bibitem[Wolter et al. (1998)]{wolter98}
Wolter, A., Comastri, A., Ghisellini, G., et al., 1998, A\&A, 335, 899.
\bibitem[Woo \& Urry (2002)]{woo02}
Woo, J.-H., Urry, C.M., 2002, AAS, 201, 1105.
\bibitem[Wozniak et al. (1998)]{wozniak98}
Wozniak, P.R., Zdziarski, A.A., Smith, D., Madejski, G.M., Johnson, W.N., 1998, MNRAS, 299, 449.
\end{thebibliography}
\end{document}